\documentclass[a4paper,11pt]{article}

\usepackage{jheppub} 

\usepackage[T1]{fontenc} 

\newcommand{\be}{\begin{equation}}
\newcommand{\ee}{\end{equation}}
\newcommand{\ba}{\begin{eqnarray}}
\newcommand{\ea}{\end{eqnarray}}

\newcommand{\n}[1]{\label{#1}}
\newcommand{\non}{\nonumber}
\newcommand{\eq}[1]{(\ref{#1})}

\newcommand{\hu}{\widehat{U}}
\newcommand{\hw}{\widehat{W}}

\newcommand{\hv}{\widehat{V}}

\newcommand{\cu}{\mathcal{U}}
\newcommand{\cw}{\mathcal{W}}
\newcommand{\cv}{\mathcal{V}}

\newcommand{\hh}{\, ,\hspace{0.5cm}}

\newcommand{\bi}[1]{\bibitem{#1}}

\newcommand{\BM}[1]{{\mbox{\boldmath $#1$}}}

\title{\boldmath Distorted Five-dimensional 
Electrically Charged Black Holes}

\author[a]{Shohreh Abdolrahimi}
\author[b]{and Andrey A. Shoom}

\affiliation[a]{Institut f\"ur Physik, 
Universit\"at Oldenburg, 
Postfach 2503 D-26111 Oldenburg, Germany}
\affiliation[b]{Theoretical Physics Institute, University of Alberta, Edmonton, Alberta T6G 2E1, Canada}

\emailAdd{shohreh.abdolrahimi@uni-oldenburg.de}
\emailAdd{ashoom@ualberta.ca}

\abstract{In this paper, we study distorted, five-dimensional, electrically charged (non-extremal) black holes on the example of a static and "axisymmetric" black hole distorted by external, electrically neutral matter. Such a black hole is represented by the derived here solution of the Einstein-Maxwell equations which admits an $\mathbb{R}^1\times U(1)\times U(1)$ isometry group. The external matter, which is "located" at the asymptotic infinity, is not included into the solution. The space-time singularities are located behind the black hole's inner (Cauchy) horizon, provided that the sources of the distortion satisfy the strong energy condition. The inner (Cauchy) horizon remains regular if the distortion fields are finite and smooth at the outer horizon. The solution has some remarkable properties. There exists a certain duality transformation between the inner and the outer horizon surfaces which links surface gravity, electrostatic potential, and space-time curvature invariants calculated at the black hole horizons. The product of the inner and outer horizon areas depends only on the black hole's electric charge and the geometric mean of the areas is the upper (lower) limit for the inner (outer) horizon area. The electromagnetic field invariant calculated at the horizons is proportional to the squared surface gravity of the horizons. The horizon areas, electrostatic potential, and surface gravity satisfy the Smarr formula. We formulated the zeroth and the first laws of mechanics and thermodynamics of the distorted black hole and found a correspondence between the global and local forms of the first law. To illustrate the effect of distortion we consider the dipole-monopole and quadrupole-quadrupole distortion fields. The relative change in the Kretschamnn scalar due to the distortion is greater at the outer horizon than at the inner one. By calculating the maximal proper time of free fall from the outer to the inner horizons we show that the distortion can noticeably change the black hole interior. The change depends on type and strength of distortion fields. In particular, due to the types of distortion fields considered here the black hole horizons can either come arbitrarily close to or move far from each other.}
\begin{document} 
\maketitle
\flushbottom

\section{Introduction}

This paper is an extension of our previous results \cite{AFS.IV} to five-dimensional space-times. In its analytical description, it is based on the paper on a distorted, five-dimensional, static, and "axisymmetric"\footnote{Here and in what follows, by "axisymmetric" we mean a five-dimensional black hole solution which admits an $U(1)\times U(1)$ isometry group.} vacuum  black hole \cite{AS-S}. 

A study of distorted black holes in four-dimensional space-times is naturally motivated by astrophysical problems. For example, a realistic black hole interacts with its accretion disk. Such an interaction affects the space-time metric, which, as a result, differs from the space-time metric of an isolated black hole. Thus, in the application to real astrophysical problems, the isolated black hole solutions, e.g., the Schwarzschild solution or the Kerr one, are highly idealized. To construct exact solutions which would model to some extend the interaction of a black hole with the external matter, Geroch and Hartle \cite{Ger.IV} proposed to consider static (or stationary), axisymmetric space-times which are not asymptotically flat. Such solutions represent black holes distorted by external matter. In the case of static and vacuum, axisymmetric space-times, the external metric near these distorted black holes is given by a Weyl solution \cite{Weyl.IV}. Such a Weyl solution represents, what is called, a local black hole solution, which is a space-time metric in the external neighborhood of a distorted black hole horizon. The external matter is not included into the solution and "located" at the asymptotic infinity. As it was demonstrated by Geroch and Hartle \cite{Ger.IV}, a local black hole solution has an extension to a "true" distorted, static black hole solution which is asymptotically flat and includes the black hole horizon and its interior, as well as the external matter. In what follows, we shall use the term "distorted black hole" for a solution which includes the black hole horizon and its interior but not asymptotically flat, i.e., it does not include the external matter, however, an extension to an asymptotically flat solution is, in principle, possible.

Four-dimensional, distorted, static, axisymmetric, vacuum black holes were studied in, e.g., \cite{IsKh.IV,MS,W1,Chandra,Fai.IV,FS1}. Such black hole solutions arise naturally in space-times where one of the spatial dimensions is compactified (see, e.g., \cite{May1,Bog,F2}). Distorted, static, axisymmetric, electrically charged black holes were studied in, e.g., \cite{W3,Fai.IV,AFS.IV}, and distorted, stationary, axisymmetric, vacuum and electrically charged black holes were studied in, e.g., \cite{Tomi,Bre.IV} and \cite{Bre2.IV,Hen1,Hen2,Hen3}, respectively. For an example of a distorted, static, axisymmetric, electrically charged dilaton black hole see \cite{Yaz0}. Distorted black holes can show some strange and remarkable properties, for example, for sufficiently distorted Schwarzschild solutions the isolated horizon can be foliated with neither marginally trapped nor outer trapping two-dimensional surfaces \cite{Pil}. As a result of distortion, the space-time curvature can become very high at some regions of black hole horizon \cite{FS1}. In this paper, we illustrate that distorted black holes have some unique and universal properties.  

Let us now discuss construction of static and axisymmetric distorted black hole solutions. In four dimensions, the general static, axisymmetric solution of the vacuum Einstein equations can be written in the form of Weyl solution which is defined by one of its metric function solving the Laplace equation in an auxiliary three-dimensional flat space. The other remaining metric function is derived by a line integral defined by the first one. Having written one of the Einstein equations as the Laplace equation
in a three-dimensional flat space is a great advantage. One can consider its solution as a Newtonian potential of some axisymmetric  source. In addition, because the Laplace equation is linear, one can use the superposition principle. As a result, distorted black hole solutions can be constructed by adding extra harmonic functions to the original one, which represents the black hole source. These functions represent the distortion fields. 

In higher-dimensional space-times we have a very rich variety of black objects classified according to their horizon topology, for example, black holes, black strings, and black rings (for a review see \cite{20}). The possibility of constructing higher-dimensional distorted black holes is based on the {\em generalized} Weyl solution described by Emparan and Reall in the remarkable paper \cite{Emp.IV}. This generalization is based on the observation that a four-dimensional Weyl solution is characterized by two orthogonal commuting Killing vector fields generating an $\mathbb{R}^1\times U(1)$ isometry group, rather than by an $\mathbb{R}^1\times O(2)$ isometry group. As a result, the generalized Weyl solution (in a $d-$dimensional space-time, $d>4$) is characterized by $d-2$ orthogonal commuting Killing vector fields. There are two classes of the generalized Weyl 
solution (see \cite{Emp.IV}). The one we shall consider here is defined by $d-3$ metric functions which solve the Laplace equation in an auxiliary three-dimensional flat space. Like in four-dimensional space-times, these functions can be considered as Newtonian potentials of certain sources. They are subject to additional constraint, which implies that the sources must add up to produce an infinite rod of zero thickness. Thus, the generalized $d-$dimensional Weyl solution is defined by $d-3$ independent axisymmetric metric functions solving the Laplace equation and the remaining metric function is derived by a line integral expressed through them. A framework for the generalized Weyl solution in five-dimensional space-times within the Einstein-Gauss-Bonnet theory was proposed by Kleihaus, Kunz, and  Radu \cite{KKR} and a numerical evidence for the existence of solutions representing a static black ring was given. The generalized Weyl solution was extended to a $d-$dimensional, $d\geqslant 4$, stationary solution characterized by $d-2$ commuting Killing vector fields by Harmark \cite{Harm}.

The generalized Weyl solution includes many interesting black objects of different horizon topology and configuration (see, e.g., \cite{20,Emp.IV}). However, only in $d=4,5$ there are solutions which can be globally asymptotically flat, for example, four- and five-dimensional Shwarzschild-Tangherlini black holes. Multi-black hole configurations, that are higher-dimensional generalizations of the Israel-Khan solutions \cite{IsKh.IV}, were discussed in \cite{Emp.IV}. Let us note that a five-dimensional solution with two black hole horizons is not asymptotically flat, while a five-dimensional solution with three black hole horizons is asymptotically flat but has irremovable conical singularities. Asymptotically flat solutions that describe a "collinear" configuration of five-dimensional Schwarzschild-Tangherlini black holes were constructed by Tan and Teo \cite{TT}. The corresponding background space-time is not flat and has conical singularities. These solutions were generalized to a configuration of charged black holes, with fixed mass-to-charge
ratio, within the Einstein-Maxwell-dilaton theory \cite{TT}. By applying a solution generation technique, which utilizes the symmetries of the reduced Lagrangian, Chng, Mann, Radu, and Stelea \cite{Chng} generated the multi-Reissner-Nordstr\"om solution in five-dimensions with general masses and electric charges from a four-dimensional multi-Reissner-Nordstr\"om solution. In that solution, the black holes distort one another by their mutual gravitational attraction. If we push the other black holes to asymptotic infinity by taking an appropriate limit in the metric and focus on the geometry near one of the black holes, we get a distorted black hole solution. However, such a solution represents a black hole distorted by a very specific configuration of external sources defined by the other black holes. We would like to find a solution representing a black hole distorted by more general configuration of external sources retaining the axial symmetry. In this paper, using another generating technique, we shall construct a solution representing a five-dimensional static electrically charged black hole distorted by "axisymmetric" configuration of electrically neutral sources. 

As it was done in the case of a four-dimensional Weyl solution, one can use the generalized Weyl solution to construct distorted black objects by adding the distortion fields to the Newtonian potentials, which define the solution. Using this approach, a distorted five-dimensional Schwarzschild-Tangherlini black hole solution was constructed and analyzed in our previous paper \cite{AS-S}. Here, we shall construct and analyze a solution representing a distorted five-dimensional Reissner-Nordstr\"om black hole. This is a static solution of the Einstein-Maxwell equations which has $\mathbb{R}^1\times U(1)\times U(1)$ isometry group. The construction is based on the gauge transformation of the matrix which is an element of the coset target space $SL(2,\mathbb{R})/U(1)$ of the scalar fields which define our model (see, e.g., \cite{GX,Maz,EGK,GK,Gal,GK1,CG,GL,GS,
GR,CGM,Yaz1,Yaz2,Yaz3}). 

Having our solution constructed, we shall analyse properties of such a black hole, focusing on its outer and inner horizons and the interior region located between them. We compare the distorted black hole solution with the Reissiner-Nordstr\"om solution representing undistorted black hole and study which properties remain and which are changed due to the distortion. The main questions we shall ask are the following:
\begin{itemize}
\item Is there a relation between geometries of the horizons similar to the one that was established for the horizons of a four-dimensional static and axisymmetric distorted black hole (see \cite{AFS.IV})? In the case of the four-dimensional black hole, such a relation is expressed through a duality transformation between its two-dimensional outer and inner horizon surfaces. An analogous relation between the Ernst and electromagnetic  potentials at the outer and inner horizons of a four-dimensional stationary and axisymmetric black hole distorted by arbitrary surrounding matter was derived by Ansorg and Hennig \cite{Hen2,Hen3}. The duality transformation is identical to the duality transformation between the horizon and the {\em stretched} singularity surfaces of a distorted four-dimensional Schwarzschild black hole \cite{FS1}. It is interesting to find out if a similar situation exists in the five-dimensional case.

\item What is the space-time curvature at the black hole horizons? In particular, does the inner (Cauchy) horizon remain regular under static and $U(1)\times U(1)-$symmetric distortions? Importance of this question is manifested in four-dimensional space-times. Namely, observers traveling inside a Reissner-Nordstr\"om or, more generally, Kerr-Newman black hole receive an infinitely blueshifted radiation when they approach the Cauchy horizon. Penrose argued that small perturbations produced in the black hole exterior grow infinitely near the Cauchy horizon \cite{Penrose}.  The evolution of small perturbations was analyzed in \cite{McN,Matz,Chands} and the derived results confirmed intuitive Penrose's arguments. Later Poisson and Israel considered incoming and outgoing radiation propagating inside an electrically charged black hole and showed that such radiation results in an infinite grow of the black hole internal mass parameter and, as a result, in an infinite grow of the space-time curvature near the Cauchy horizon \cite{PI}. An exact and simplified solution describing this phenomenon was constructed by Ori \cite{Ori}. For more details about instability of the Chauchy horizon see \cite{FroNo} and references therein.  Maeda, Torii, and  Harada have demonstrated a novel the so-called kink instability of the Cauchy horizon of $d-$dimensional static solutions in the Einstein-Maxwell-scalar-$\Lambda$ system, which includes the Reissner-Nordstr\"om-(anti-) de Sitter
black hole \cite{Maeda}. In our previous work \cite{AFS.IV}, we have shown that the inner (Cauchy) horizon of a distorted, four-dimensional, electrically charged, static, and axisymmetric black hole is regular, provided the distortion field is regular and smooth at its outer horizon. Thus, it is natural to ask if that is true in the five-dimensional case.  

\item What is the effect of distortion on the black hole interior? In particular, what is the effect of distortion on the maximal proper time of free fall from the outer to the inner horizon of the black hole? In the case of a distorted, four-dimensional, electrically charged, static, and axisymmetric black hole it was showed that the distortion noticeable changes the maximal proper time of free fall of a test particle moving in the black hole interior along the symmetry axis. Here we shall study the same problem in the case of a five-dimensional distorted black hole. 

\item There are other interesting issues that can be addressed here. Namely, the product of the black hole horizon areas and the areas inequality. The horizon area product of a four-dimensional, stationary and axisymmetric, electrically charged black hole distorted by arbitrary surrounding matter distribution was derived by Ansorg and Hennig \cite{Hen1,Hen2,Hen3}. The product is expressed in terms of the black hole's angular momentum and electric charge. They showed that the horizon areas of such a black hole satisfy an inequality, which defines the upper and the lower limits on the values of its inner and outer horizon areas, respectively (see \cite{Hennig1} as well). Cveti\v c, Gibbons, and Pope \cite{Cvetic1} have shown that the product of all horizon areas of a general rotating multicharge black hole in asymptotically flat or asymptotically anti-de Sitter space-times of four and higher dimensions depends only on its charges, angular momenta, and the cosmological constant, and it is independent of its mass. The product is quantized within the framework of a weakly-coupled two-dimensional conformal field theory. Visser \cite{Vis} has shown that generically, products of horizon areas may not be independent of the black hole's mass. For example, the product of the Schwarzschild-de Sitter (Kottler) black hole horizon area and the area of the cosmological horizon depends on the black hole mass. The product of  Reissner-Nordstr\"om-anti-de Sitter black hole horizon areas depends on mass as well. In this work, we shall analyse the horizon areas product and inequality of the distorted, five-dimensional, electrically charged black hole. 
\end{itemize} 
 
Our paper is organized as follows. In Sec.~2 we present the five-dimensional Reissner-Nordstr\"om black hole in the suitable coordinates and calculate its horizon areas, surface gravity, and electrostatic potential values at its horizons. We derive the Smarr formula for both the black hole horizons. In Sec.~3 we present a transformation which produces a five-dimensional, electrically charged, static solution to the Einstein-Maxwell equations, when applied to a five-dimensional, vacuum, static seed solution. We illustrate the transformation on the example of five-dimensional Schwarzschild-Tangherlini and Reissner-Nordstr\"om black holes. In Sec.~4 we apply the transformation to a solution representing distorted, five-dimensional, vacuum, static  black hole given by the generalized Weyl form and construct the solution representing five-dimensional, electrically charged, static black hole distorted by electrically neutral external matter distribution. We give answers to the questions raised above in the following sections. Namely, in Sec.~5 we present the duality transformation between the black hole's outer and inner horizons. We  calculate its horizon areas, surface gravity, and electrostatic potential values at its horizons. Using these results we calculate the horizon areas product, define the area inequality, and derive the Smarr formula for both the black hole horizons. In Sec.~6 we discuss mechanics and thermodynamics of the distorted black hole. In particular, we formulate the zeroth law and the global and local form of the first law for the black hole outer and inner horizons. We show the correspondence between the global and local forms.  In Sec.~7 we study the space-time curvature at the black hole horizons. We derive the Kretschmann scalar and show that its values at the black hole horizons are related by the duality transformation. In Sec.~8 we study the model of a five-dimensional, electrically charged, static and "axisymmetric" black hole distorted  adiabatically by external, electrically neutral matter. In such a model the black hole horizon areas do not change and are equal to those of an undistorted (Reissner-Nordstr\"om) black hole of the given mass and charge. This model allows us to study the effect of distortion on the black hole horizons and interior by varying the parameters of the distortion fields. We consider an example of the dipole-monopole and quadrupole-quadrupole distortions and calculate the corresponding values of the Kretschmann scalar at the black hole horizons and the maximal proper time of free fall of a test particle from the outer to the inner horizon along the symmetry "semi-axes". Section~9 of our paper contains the summary and discussion of the derived results.
 
In this paper we use the following convention of units: $G_{(5)}=c=\hbar=k_{B}=1$, where $G_{(5)}$ is the five-dimensional gravitational constant. The space-time signature is +3 and the sign conventions are that adopted in \cite{MTW}.

\section{The five-dimensional Reissner-Nordstr\"om black hole}

The Einstein-Maxwell theory in five-dimensions is described by the action
\be\n{2.1}
S=\frac{1}{16\pi}\int d^5x \sqrt{-g}
\left(R-\frac{1}{4}F^2\right)\,,
\ee
where $g=\text{det}(g_{\mu\nu})$ is the determinant of the five-dimensional space-time metric $g_{\mu\nu}$, $R$ is the five-dimensional Ricci scalar, and $F^2=F_{\mu\nu}F^{\mu\nu}$, where $F_{\mu\nu}=\nabla_{\mu}A_{\nu}
-\nabla_{\nu}A_{\mu}$ is the electromagnetic field tensor. 

The Einstein-Maxwell field equations derived from this action are
\ba
&&R_{\mu\nu}-\frac{1}{2}g_{\mu\nu}R=8\pi T_{\mu\nu}\hh 8\pi T_{\mu\nu}=\frac{1}{2}\left(F_{\mu}^{~\lambda}F_{\nu\lambda}-\frac{1}{4}g_{\mu\nu}F^2\right)\,,\n{2.2}\\
&&\hspace{4cm}\nabla_{\lambda}F^{\mu\nu}=0\,.\n{2.3}
\ea
Here and in what follows, $\nabla_\mu$ denotes the covariant derivative defined with respect to the metric $g_{\mu\nu}$.

The five-dimensional Reissner-Nordstr\"om solution is a static, spherically symmetric, asymptotically flat solution of the Einstein-Maxwell equations (see, e.g., \cite{Tang,MP}) that corresponds to the five-vector potential
\be\n{2.4}
A_{\mu}=-\Phi\delta_{\mu}^{t}\,,
\ee
where $\Phi=\Phi(r)$ is an electrostatic potential. The solution reads 
\ba
ds^2&=&-fdt^2+f^{-1}dr^2+{r^2}d\Omega^2\hh f=1-\frac{2m}{r^2}+\frac{q^2}{r^4}\,,\n{2.5}\\
\Phi&=&\frac{\sqrt{3}q}{r^2}\hh F_{\mu\nu}=-\frac{2\sqrt{3}q}{r^{3}}(\delta_\mu^t\delta_\nu^r-
\delta_\mu^r\delta_\nu^t)\,,\n{2.6}
\ea
where $m$ and $q$ are parameters of the solution, $d\Omega^2$ is a metric on a round unit three-sphere, which can be presented in the following form:
\be\n{2.7}
d\Omega^2=\frac{1}{4}\left(d\theta^2+2(1+\cos\theta)d\chi^2
+2(1-\cos\theta)d\phi^2\right)\,,
\ee
where $\theta\in[0,\pi]$, $\chi\in[0,2\pi)$, and $\phi\in[0,2\pi)$ are the Hopf coordinates. The parameters $m$ and $q$ are related to the five-dimensional Komar mass of the black hole $M$ and its five-dimensional electric charge $Q$  as follows:
\be\n{2.8}
M=\frac{3\pi}{4}m\hh Q=2\sqrt{3}\,q\,.
\ee
The five-dimensional Komar mass of the black hole defined by the following expression:
\be\n{2.9}
M=-\frac{3}{32\pi}\oint_{S_{\infty}} d^{3}\Sigma_{\mu\nu}\nabla^\mu\xi^\nu\,,
\ee
where $\xi^\mu_{(t)}=\delta^\mu_t$ is a time-like Killing vector normalized at the spatial infinity, $\BM{\xi}^2_{(t)}=-1$, 
\be\n{2.10}
d^{3}\Sigma_{\mu\nu}=\frac{1}{3!}\sqrt{-g}\,\varepsilon_{\mu\nu\lambda\sigma\rho}\,
dx^{\lambda}\wedge dx^{\sigma}\wedge dx^{\rho}\hh \varepsilon_{tr\theta\chi\phi}=+1\,,
\ee
is the area element of a three-dimensional closed space-like surface at the spatial infinity, $S_{\infty}$. The five-dimensional electric charge $Q$ is defined as follows:
\be\n{2.11}
Q=\frac{1}{4\pi^2}\oint_{S}d^{3}
\Sigma_{\mu\nu}F^{\mu\nu}\,,
\ee
where $S$ is a three-dimensional closed space-like surface. 

The black hole horizons are at
\be\n{2.11a}
r_{\pm}=(m\pm \sqrt{m^2-q^2})^{\frac{1}{2}}\,,
\ee
where the upper sign stands for the event (outer) horizon and the lower sign stands for the Cauchy (inner) horizon. To indicate that a quantity (...) is calculated at the black hole horizons we shall use the subscripts $\pm$ and denote such a quantity as $(...)_{\pm}$. Accordingly, in the expressions what follow, the upper sign stands for the outer horizon and the lower one stands for the inner horizon, unless stated otherwise. The space-time \eq{2.5} has a time-like singularity at $r=0$. In what follows, we shall consider non-extremal black holes with $q<m$  and without the loss of generality take $q> 0$.

For our future purposes it is convenient to introduce the following coordinate transformation\footnote{Note that this coordinate transformation is valid for a non-extremal black hole only, i.e., for $p>0$.}:
\be\n{2.12}
r=\sqrt{m(1+p\eta)}\hh p=\frac{1}{m}\sqrt{m^2-q^2}\,,
\ee
where $p\in(0,1)$ and $\eta\in(-1/p,+\infty)$. Then, the black hole horizons are at
\be\n{2.12a}
r_{\pm}=\sqrt{m(1\pm p)}\,.
\ee
Using the transformation, we can rewrite the solution \eq{2.5}--\eq{2.6} in the form
\ba
&&ds^2=-\frac{p^2(\eta^2-1)}{(1+p\eta)^2}dt^2+m(1+p\eta)\left(\frac{d\eta^2}{4(\eta^2-1)}+d\Omega^2\right)\,,\n{2.13}\\
&&\Phi=\frac{\sqrt{3(1-p^2)}}{1+p\eta}\hh F_{\mu\nu}=-\frac{p\sqrt{3(1-p^2)}}{(1+p\eta)^2}(\delta_\mu^t\delta_\nu^\eta-
\delta_\mu^\eta\delta_\nu^t)\,,\n{2.14}
\ea
where $d\Omega^2$ is given by Eq.\eq{2.7}. In these coordinates the event horizon is at $\eta=1$ and the Cauchy horizon is at $\eta=-1$, and the space-time singularity is at $\eta=-1/p$.

The horizons surface areas are
\be\n{2.15}
\mathcal{A}_\pm=2\pi^2
\sqrt{m^3(1\pm p)^3}\,.
\ee
From this expression we derive the horizon areas product
\be\n{2.15a}
\mathcal{A}_+\mathcal{A}_-=4\pi^4q^3
=\frac{\pi^4}{6\sqrt{3}}Q^3\,.
\ee
The horizon areas product for distorted four-dimensional axisymmetric stationary black holes was derived by Ansorg and Hennig \cite{Hen1,Hen2,Hen3}. Such a product is quite a universal result valid for many charged and rotating black hole solutions in four- and higher-dimensional asymptotically flat or anti-de Sitter spacetimes (see, e.g., \cite{Cvetic1,Vis,Rod}).
 
The surface gravity at the horizons is
\be\n{2.16}
\kappa^2_\pm=-\frac{1}{2}(\nabla^\mu \xi^\nu)(\nabla_\mu \xi_\nu)|_\pm=\frac{4p^2}{m(1\pm p)^3}\hh \kappa_\pm=\frac{2p}{\sqrt{m(1\pm p)^3}}\,,
\ee
and the electrostatic potential at the horizons is
\be\n{2.16a}
\Phi_{\pm}=\frac{\sqrt{3(1-p^2)}}{1\pm p}\,.
\ee
Using these expressions we can construct the Smarr formula of the black hole horizons,
\be\n{2.17}
\pm M=\frac{3}{16\pi}\kappa_\pm \mathcal{A}_\pm\pm\frac{\pi}{8}\Phi_\pm Q\,.
\ee
The change of signs in the Smarr formula for the Cauchy horizon is due to space-like nature of the Killing vector $\xi^\mu_{(t)}$ in the region between the horizons, that implies negative energy. The Smarr formula for the inner horizon of asymptotically flat and anti-de Sitter higher-dimensional solutions was given in, e.g., \cite{Rod}. 

Calculating the electromagnetic field invariant $F^2$ at the black hole horizons we find the following relation:
\be\n{2.18}
F^2_\pm=-\frac{6}{p^2}(1-p^2)\kappa^2_\pm\,.
\ee
   
In the case of $q=0$, i.e., for $p=1$, the electrostatic potential $\Phi$ vanishes, and the metric \eq{2.13} represents the vacuum five-dimensional Schwarszchild-Tangherlini space-time \cite{Tang},
\ba\n{2.19}
ds^2&=&-\frac{\eta-1}{\eta+1}dt^2+m(\eta+1)\left(\frac{d\eta^2}{4(\eta^2-1)}+d\Omega^2\right)\,.
\ea
Note that according to the transformation \eq{2.12}, in this case $r=\sqrt{m(\eta+1)}$, where $\eta
\in(-1,+\infty)$.

\section{Charging vacuum solutions}

In this section, we present a transformation which, when applied to a five-dimensional static vacuum seed solution, produces a charged static solution of the Einstein-Maxwell equations. 
We shall follow the procedure based on the gauge transformation of the matrix which is an element of the coset target space $SL(2,\mathbb{R})/U(1)$ of the scalar fields which define our model (see, e.g., \cite{GX,Maz,EGK,GK,Gal,GK1,CG,GL,GS,
GR,CGM,Yaz1,Yaz2,Yaz3}).

\subsection{The generating transformation}

The metric of a five-dimensional static spacetime can be written in the following form:
\be\n{3.1}
ds^2=-e^{2U}dt^2+e^{-U}h_{ij}dx^idx^j\,,
\ee
where $U=U(x^{i})$, $h_{ij}=h_{ij}(x^i)$,  $i,j=(1,2,3,4)$. Using this form of the metric we can present the quantities in the action \eq{2.1} in terms of the four-dimensional metric $h_{ij}$ and the metric function $U$ as follows:
\be\n{3.2}
\sqrt{-g}=e^{-U}\sqrt{h}\hh R=e^{U}\left(\mathcal{R}-\frac{3}{2}h^{ij}U_{,i}U_{,j}+\triangle U\right)\hh 
F^2=-2e^{-U}h^{ij}\Phi_{,i}\Phi_{,j}\,,
\ee
where $h=\text{det}(h_{ij})$, $\mathcal{R}$ is the four-dimensional Ricci scalar defined by the metric $h_{ij}$, $\triangle$ is the Laplace-Beltrami operator defined with respect to the metric $h_{ij}$. Here and in what follows, the expression $(...)_{,i}$ means the partial derivative of $(...)$ with respect to the coordinate $x^i$. We derived the expression for the Ricci scalar $R$ by using a dimension reduction along the Killing coordinate $t$, followed by a conformal transformation applied to the reduced four-dimensional Ricci scalar.

Substituting these expressions into the action \eq{2.1}, integrating by parts, and neglecting the surface terms we derive the following equivalent action:
\ba\n{3.3}
\mathcal{S}=\frac{1}{16\pi}\int d^4x \sqrt{h}\left(\mathcal{R}-\frac{3}{2} {h}^{ij} U_{,i}U_{,j}+\frac{1}{2}e^{-2U}h^{ij}\Phi_{,i}
\Phi_{,j}\right)\,.
\ea 
This action represents four-dimensional nonlinear $\sigma-$model without dilaton field. Following \cite{Gal,GR}, one can define the target space metric defined by the scalar fields $U$ and $\Phi$ as
\be\n{3.3a}
dl^2=\frac{3}{2}dU^2-\frac{1}{2}e^{-2U}d\Phi^2\,,
\ee
which represents a coset $SL(2,\mathbb{R})/U(1)$, where $SL(2,\mathbb{R})$ is an isometry group acting on the target space and $U(1)$ is its isotropy subgroup. To construct the matrix corresponding to the target space we follow the procedure presented in, e.g. \cite{Maz,EGK,Gal}, that gives 
\be\n{3.5}
A=\begin{pmatrix} e^{U}-\frac{\Phi^{2}}{{3}}e^{-U}&&&&-\frac{\Phi}{\sqrt{3}}e^{-U}\\
-\frac{\Phi}{\sqrt{3}}e^{-U}&&&&-e^{-U}\end{pmatrix}\in SL(2,\mathbb{R})/U(1)\,,
\ee
which allows for the following matrix representation of the $\sigma-$model:
\be\n{3.4}
\mathcal{S}=\frac{1}{16\pi}\int d^4x \sqrt{h}\left(\mathcal{R}+\frac{3}{4} {h}^{ij}\text{Tr}\left\{A_{,i} (A^{-1})_{,j}\right\}\right)\,.
\ee
One can easily check that the action \eq{3.4}, when extremized, generates the Einstein-Maxwell equations,  
\be\n{3.6}
\triangle U=\frac{1}{3}e^{-2U}h^{ij}
\Phi_{,i}\Phi_{,j}\hh\mathcal{R}_{ij}
=\frac{3}{2}U_{,i}U_{,j}
-\frac{1}{2}e^{-2U}\Phi_{,i}\Phi_{,j}\hh
\left(e^{-2U}h^{ij}\Phi_{,i}\right)_{,j}
=0\,,
\ee
written in terms of the static metric functions \eq{3.1} and the electrostatic potential $\Phi$. The action \eq{3.4} is invariant under the symmetry transformation (see, e.g., \cite{Yaz1}) 
\be\n{3.7}
\bar{A}=GA\,G^T\,,
\ee
where $G\in SL(2,\mathbb{R})$ is a constant matrix. 

To consider a specific form of the transformation matrix $G$, we need to specify the matrix $A$ in order that its properties will be preserved under the transformation \eq{3.7}. Let us consider asymptotically flat space-times. In such space-times, one can find an appropriate coordinate system in which the metric function $U$ and the electrostatic potential $\Phi$ have asymptotic expansions of the form
\be\n{3.8}
e^{2U}=1-\frac{C_1}{r^{2}}+{\cal O}(1/r^{3})\hh \Phi=\frac{C_2}{r^{2}}+{\cal O}(1/r^{3})\,,
\ee
where $r$ is the radial distance, the constant $C_1$ is proportional to the Komar mass \eq{2.9}, and the constant $C_2$ is proportional to the electric charge \eq{2.11}. For an asymptotically flat solution the matrix $A$ at infinity is 
\be\n{3.9}
A_\infty=\begin{pmatrix} 1&&&0\\
0&&&-1\end{pmatrix}\,.
\ee
This matrix can be considered as a matrix representing the metric of two-dimensional Minkowski space-time. Thus, to preserve the asymptotic flatness\footnote{For other class of transformations which do not preserve asymptotic flatness see, e.g., \cite{Yaz1,Yaz2,Yaz3}.}, $G$ should satisfy the condition
\be
G\,A_\infty G^T=A_\infty\,.
\ee
This implies that one can choose
\be\n{3.11}
G=\begin{pmatrix} \cosh\delta &&&\sinh\delta\\
\sinh\delta &&&\cosh\delta\end{pmatrix}\in SO(1,1)\subset SL(2,\mathbb{R})\,,
\ee
where $\delta\in \mathbb{R}^1$ is the "boost" parameter.

To illustrate this transformation let us consider a static asymptotically flat vacuum space-time of the form \eq{3.1}, then
\be\n{3.12} 
A=\begin{pmatrix} e^{U}&&&0\\ 
0&&&-e^{-U}\end{pmatrix}\,.
\ee
Applying the transformation \eq{3.7} defined by \eq{3.11} we derive
\ba
ds^2&=&-e^{2\bar{U}}dt^2+e^{-\bar{U}}
h_{ij}dx^idx^j\,,\n{3.13}\\
e^{\bar{U}}&=&\frac{e^{U}}{(\cosh^2\delta-e^{2U}\sinh^2\delta)}\,,\n{3.14}\\
\bar{\Phi}&=&-\sqrt{3}
\frac{(e^{2U}-1)\tanh\delta}{(1-e^{2U}\tanh^2\delta)}\,.\n{3.15}
\ea
This is a static solution of the five-dimensional Einstein-Maxwell equations \eq{3.6}.\footnote{Note, that the transformation \eq{3.14}--\eq{3.15} can be derived by introducing the Ernst potentials and applying to them the Harrison-Ernst transformation with $c=\tanh\delta$, followed by gauge transformation with $b=-\sinh\delta\cosh\delta$ and scaling transformation with $a=\cosh^{-2}\delta$ ( see Eqs.(26)--(28), \cite{GR}).} We call the above transformation \eq{3.14}--\eq{3.15} the generating transformation. In what follows, we shall apply this transformation to a five-dimensional Weyl solution. 

\subsection{The Weyl seed solution}
 
A five-dimensional Weyl solution is characterized by three commuting, orthogonal Killing vector fields, $\xi^\mu_{(t)}=\delta^\mu_{t}$, $\xi^\mu_{(\chi)}=\delta^\mu_{\chi}$, and $\xi^\mu_{(\phi)}=\delta^\mu_{\phi}$ (for details see \cite{Emp.IV}). A five-dimensional Weyl solution can be presented as follows:
\ba\n{3.16}
ds^2&=&-e^{2U_1}dt^2+e^{2V}(\eta^2-\cos^2\theta)\left(\frac{d\eta^2}{\eta^2-1}+d\theta^2\right)
+e^{2U_2}d\chi^2+e^{2U_3}d\phi^2\,,
\ea
where $\theta\in[0,\pi]$, $\chi\in[0,2\pi)$, and $\phi\in[0,2\pi)$ are the Hopf coordinates [see Eq.\eq{2.7}]. The metric functions $U_i$, $i=1,2,3$, and $V$ depend on the coordinates $\eta$ and $\theta$. Each of the functions $U_i$ solves the following three-dimensional Laplace equation:
\be\n{3.17}
(\eta^2-1)U_{i,\eta\eta}+2\eta\,U_{i,\eta}+U_{i,\theta\theta}+\cot\theta\,U_{i,\theta}=0\,,
\ee
and the following constraint holds:
\be\n{3.18}
e^{2(U_1+U_2+U_3)}=(\eta^2-1)\sin^2\theta.
\ee
Once the functions $U_i$'s are found, the metric function $V$ can be calculated from a line integral involving the functions $U_i$'s and their first-order derivatives (for details see, e.g., \cite{Emp.IV,20}, and \cite{AS-S}). 

The metric \eq{3.16} is a solution of vacuum Einstein's equations. Let us denote $U=U_1$ and introduce the conformal factor $e^{-U_1}$ in the spatial part of the metric \eq{3.16}, then we can identify this metric  with the metric \eq{3.1}. Applying the generating transformation \eq{3.14}--\eq{3.15} to the Weyl metric one can derive a solution of the Einstein-Maxwell equations \eq{3.6}. In what follows, it is convenient to introduce a parameter $p$ which is related to the "boost" parameter $\delta$,
\be\n{3.19}
\cosh\delta=\sqrt{\frac{1+p}{2p}}\hh \sinh\delta=\pm\sqrt{\frac{1-p}{2p}}\hh 0<p\leqslant 1\,.
\ee
Here we shall restrict ourselves to positive values of $\delta$ and $p\in(0,1)$. As we shall see below, such a restriction corresponds to non-extremal Reissner-Nordstr\"om solution with positive electric charge and $p$ given by Eq.\eq{2.12}. Then, the solution generated from the Weyl metric reads  
\be\n{3.20}
ds^2=-e^{2\bar{U}_1}dt^2+e^{-\bar{U}_1+U_1}
\left(e^{2V}(\eta^2-\cos^2\theta)\left(\frac{d\eta^2}{\eta^2-1}+d\theta^2\right)
+e^{2U_2}d\chi^2+e^{2U_3}d\phi^2\right)\,,
\ee
and the generating transformation \eq{3.14}--\eq{3.15} takes the form
\ba
e^{\bar{U}_1}&=&\frac{2p\,e^{U_1}}{(1+p-(1-p)e^{2U_1})}\,,\n{3.21}\\
\Phi&=&\frac{\sqrt{3(1-p^2)}(1-e^{2U_1})}{(1+p-(1-p)e^{2U_1})}\,,\n{3.22}
\ea
where we dropped the bar from the electrostatic potential $\Phi$. Note that this transformation is almost identical to the Harrison-Ernst transformation \cite{Har,Ern} of a four-dimensional spacetime (see, e.g., \cite{AFS.IV}). It differs only by the factor $\sqrt{3}$ in the expression for $\Phi$.

As an illustrative example, let us apply the transformation \eq{3.21}--\eq{3.22} to the spacetime \eq{2.19} representing the Schwarzschild-Tangherlini black hole. By identifying the metric \eq{2.19} with the Weyl solution \eq{3.16} we find 
\ba\n{3.23}
e^{2{U}_1}&=&\frac{\eta-1}{\eta+1}\hh
e^{2{U}_2}=\frac{m}{2}(\eta+1)(1+\cos\theta)\,,\non\\
e^{2{U}_3}&=&\frac{m}{2}(\eta+1)(1-\cos\theta)\hh
e^{2V}=\frac{m(\eta+1)}{4(\eta^2-\cos^2\theta)}\,.
\ea
Then, using the transformation \eq{3.21}--\eq{3.22} we construct the metric \eq{3.20}, which takes the following form:
\be\n{3.24}
ds^2=-\frac{p^2(\eta^2-1)}{(1+p\eta)^2}dt^2+\frac{m}{p}(1+p\eta)\left(\frac{d\eta^2}{4(\eta^2-1)}+d\Omega^2\right)\,,
\ee
where $d\Omega^2$ is given by Eq.\eq{2.7} and $p$ is given by Eq.\eq{2.12}. The electrostatic potential $\Phi$ and the corresponding electromagnetic field tensor $F_{\mu\nu}$ are given by Eq.\eq{2.14}. 

Let us compare this solution with the five-dimensional Reissner-Nordstr\"om space-time \eq{2.13}. We see that the solutions match, except for the constant conformal factor $1/p$ in the spatial part of the generated metric.\footnote{Note that if we applied the transformation to the ``standard Schwarzschild form'' (see Eqs.\eq{2.5}--\eq{2.6} with $Q=0$), we would derive the Reissner-Nordstr\"om solution of the form different from that given by the expression \eq{2.5}--\eq{2.6}. In order to bring this solution to the form \eq{2.5}--\eq{2.6} one has to construct an involved coordinate transformation.} This factor can be removed by rescaling the $t$ coordinate and by subsequent rescaling of the metric as follows:
\be\n{3.25}
t\longrightarrow \frac{t}{\sqrt{p}}\hh ds^2\longrightarrow p\,ds^2\,.
\ee

\section{Distorted five-dimensional electrically charged black hole}

In this section, we apply the generating  transformation presented in the previous section to a five-dimensional Schwarzschild-Tangherlini black hole distorted by static and neutral distribution of external matter \cite{AS-S}. The distorted five-dimensional vacuum black hole solution is not asymptotically flat. Therefore, the sought distorted five-dimensional electrically charged black hole solution is not expected to be asymptotically flat either. In principle, one can consider more general transformation, which does not produce an asymptotically flat solution when applied to an asymptotically flat one (see, e.g., \cite{Yaz1,Yaz2,Yaz3}). Here we would like to construct an electrically charged distorted black hole solution, which is not asymptotically flat due to distorting matter only, in a way similar to our earlier work \cite{AFS.IV}, such that when the distortion fields vanish the space-time becomes asymptotically flat. Thus, to achieve our goal, we shall consider the transformation \eq{3.21}--\eq{3.22}.

\subsection{The vacuum seed solution}

Let us present the distorted five-dimensional Schwarzschild-Tangherlini solution in the form \eq{3.1} suitable for the transformation (for details see \cite{AS-S}),
\ba
&&\hspace{2cm}ds^2=-e^{2U}dt^2+e^{-U}h_{ij}dx^idx^j\hh e^U=\sqrt{\frac{\eta-1}{\eta+1}}e^{\hu+\hw}\,,\non\\ 
&&\hspace{3.5cm}h_{ij}dx^idx^j=\frac{m}{4}
\sqrt{\eta^2-1}e^{\hu+\hw}dl^2\,,\n{4.1}\\
&&\hspace{0cm}dl^2=e^{2(\hv+\hu+\hw)}
\left(\frac{d\eta^2}{\eta^2-1}+d\theta^2\right)
+2(1+\cos\theta)e^{-2\hw}d\chi^2
+2(1-\cos\theta)e^{-2\hu}d\phi^2\,.\non
\ea
Here $\hu$, $\hw$, and $\hv$ are the distortion fields given by the following expressions: 
\ba
\hu(\eta,\theta)&=&\sum_{n\geq0}a_n\,
R^{n}P_n\hh\hw(\eta,\theta)=\sum_{n\geq0}b_n\,
R^{n}P_n\hh\hv =\hv_1+\hv_2\,,
\n{4.2}\\
\hv_1(\eta,\theta)&=&-\sum_{n\geq0}
\Big\{3(a_n/2+b_n/2)R^nP_n
+(a_n+b_n/2)\sum_{l=0}^{n-1}(\eta-\cos\theta)R^lP_l\non\\
&+&(a_n/2+b_n)\sum_{l=0}^{n-1}(-1)^{n-l}(\eta+\cos\theta)R^lP_l\Big\}\,,\n{4.3}\\
\hv_2(\eta,\theta)&=&\sum_{n,k\geq1}\frac{nk}{n+k}(a_na_k+a_nb_k+b_nb_k)R^{n+k}
[P_nP_k-P_{n-1}P_{k-1}]\,,\n{4.4}
\ea
where
\be\n{4.5}
R=(\eta^2-\sin^2\theta)^{1/2}
\hh P_n\equiv P_n(\eta\cos\theta/R)\,.
\ee
Here $P_{n}$'s  are the Legendre polynomials of the first kind. These series generally converge if the sources of the distortion fields are located far from the black hole and the fields are considered in the black hole vicinity. Accordingly, as in the Newtonian gravitational theory and electromagnetism, the constant coefficients $a_{n}$'s and $b_{n}$'s, which define the distortion fields, are called {\em interior multipole moments}.\footnote{Note that despite the fact that the equation for the distortion fields $\hu$ and $\hw$ is Laplace one, these fields are relativistic. To construct the corresponding Newtonian fields one has to take the nonrelativistic limit, e.g., $\lim_{c^2\to\infty}c^2\,\hu(\eta,\theta;c^2)$, where $c$ is the speed of light (see, e.g., \cite{Ehlers,Quevedo}). 
Distortion fields defined by exterior multipole moments correspond to asymptotically flat solutions. However, in this case the series representing such distortion fields near the black hole horizon converge if the sources of the fields are located inside the black hole.} On the other hand, according to the uniqueness theorem formulated in \cite{57}, a Schwarzschild-Tangherlini black hole is the only $d-$dimensional, asymptotically flat, static, vacuum black hole which has non-degenerate regular event horizon. This implies that the sources located inside the black hole make its horizon singular. Thus, to have a regular horizon we shall consider non-asymptotically flat solution distorted by the external sources only, whose distortion fields are defined by the interior multipole moments. Such fields must be regular and smooth at the horizon. The distortion fields $\hu$, $\hw$, and $\hv$ given above satisfy this condition. Note that the distortion field $\hv$ is defined up to an additive constant of integration. The form of the distortion field $\hv$ corresponds to a particular choice of that constant (see footnote on page 15). There is an additional restriction on the multipole moments, which follows from the strong energy condition imposed on the distortion fields due to the positive mass theorem \cite{GH}. If the sources of the distortion fields are included into the solution, then their energy-momentum tensor $T^{(\text{source})}_{\mu\nu}$ satisfies the strong energy condition,
\be\n{4.10}
T^{(\text{source})}_{\mu\nu}-\frac{g_{\mu\nu}}{3}g_{\mu\nu}T^{(\text{source})}\geqslant 0\hh T^{(\text{source})}=g^{\mu\nu}
T^{(\text{source})}_{\mu\nu}\,.
\ee
Using the Einstein equations we derive
\be
\left(T^{(\text{source})}_{\mu\nu}-\frac{g_{\mu\nu}}{3}g_{\mu\nu}T^{(\text{source})}\right)\delta^\mu_{\,\,t}\delta^\nu_{\,\,t}=\frac{m(\eta-1)(\eta^2-\cos^2\theta)}{8\pi(\eta+1)^2}e^{-2\hv}\triangle(\hu+\hw)\,.
\ee
Because the Laplace operator $\triangle$ is negative, the strong energy condition implies that
\be\n{4.11} 
\hu+\hw\leqslant 0\,.
\ee
Then from Eq.\eq{4.2} it follows that on the "semi-axes" $\theta=0$ and $\theta=\pi$\footnote{We call the spatial regions $\theta=0$ and $\theta=\pi$ "semi-axes" in analogy with a three-dimensional space. However, in a four-dimensional space these regions are two-dimensional planes, $(\eta,\chi)$ and $(\eta,\phi)$.}, at the black hole horizon $\eta=1$,
we have
\be\n{4.12} 
\sum_{n\geq0}(\pm 1)^n(a_{n}+b_{n})\leqslant 0\,,
\ee
where $+1$ corresponds to $\theta=0$ and $-1$ corresponds to $\theta=\pi$.
 
\subsection{The solution} 

Applying the generating transformation \eq{3.21}--\eq{3.22} to the metric \eq{4.1} accompanied by the scaling transformation \eq{3.25} we derive the solution representing distorted charged black hole,
\ba
ds^2&=&-\frac{4p^2(\eta^2-1)}{\Delta^2}e^{2(\hu+\hw)}dt^2+\frac{m\Delta}{2}\left(e^{2(\hv+\hu+\hw)}\frac{d\eta^2}{4(\eta^2-1)}+d\hat{\Omega}^2\right)
\,,\n{4.13}\\
\Delta&=&(1+p)(\eta+1)-(1-p)(\eta-1)e^{2(\hu+\hw)}\,,\n{4.14}\\
d\hat{\Omega}^2&=&\frac{1}{4}\left(e^{2(\hv+\hu+\hw)}d\theta^2
+2(1+\cos\theta)e^{-2\hw}d\chi^2
+2(1-\cos\theta)e^{-2\hu}d\phi^2\right)\,,
\n{4.15}\\
\Phi&=&\frac{\sqrt{3(1-p^2)}}{\Delta}\left(\eta+1-
(\eta-1)e^{2(\hu+\hw)}\right)\,.\n{4.16}
\ea
The only non-zero components of the electromagnetic field tensor $F_{\mu\nu}$ are 
\ba
F_{t\eta}&=&-\frac{4p\sqrt{3(1-p^2)}}{\Delta^{2}}e^{2(\hu+\hw)}[1+(\eta^2-1)(\hu_{,\eta}+\hw_{,\eta})]\,,\n{4.17}\\
F_{t\theta}&=&-\frac{4p\sqrt{3(1-p^2)}}{\Delta^{2}}e^{2(\hu+\hw)}(\eta^2-1)
[\hu_{,\theta}+\hw_{,\theta}]\,.\n{4.18}
\ea
For $p=1$, this solution represents the distorted five-dimensional Schwarzschild-Tangherlini black hole \eq{4.1}. 
If the distortion fields vanish, this solution represents the five-dimensional Reissner-Nordstr\"om solution \eq{2.13}--\eq{2.14}. Note that the distorted black hole electric charge $Q$ is independent of the values of the multiple moments and equals to that of the Reissner-Nordstr\"om black hole [see Eq.\eq{2.8}].
The transformation does not affect the distortion fields $\hu$, $\hw$, and $\hv$. Thus, the transformation electrically charges the black hole only.\footnote{For example, if we consider a triple Israel-Khan black hole solution, which represents the central black hole distorted by the other two, one can show that the Ernst-Harrison transformation charges the central black hole.}   

We were able to construct a solution which represents a five-dimensional charged black hole distorted by external electrically neutral sources, such that when the distortion fields $\hu$, $\hw$, and $\hv$ vanish, the solution represents a five-dimensional Reissner-Nordstr\"om solution in an empty, asymptotically flat universe. The metric \eq{4.13}--\eq{4.15} is not asymptotically flat. The uniqueness of static, electrically charged, and  asymptotically flat black holes proved in \cite{Gib2} implies that if instead of the interior multiple moments corresponding to the external sources we considered exterior multiple moments or both, then the resulting solution would have a singular horizon.  

The distorted black hole solution \eq{4.13}--\eq{4.16} possesses two horizons, the outer horizon, at $\eta=1$, and the inner horizon, at $\eta=-1$. One can show that the space-time curvature invariants, e.g., the Kretschmann scalar, and the electromagnetic field invariant $F^2$  diverge in the region where $\Delta=0$ [see Eq.\eq{4.14}], i.e., where 
\be\n{4.19}
\eta+\frac{1+p+(1-p)e^{2(\hu+\hw)}}{1+p-(1-p)e^{2(\hu+\hw)}}=0\,.
\ee
This expression implies that the space-time singularities can be located behind the inner horizon, if there $\hu+\hw<1/2\,\ln[(1+p)/(1-p)]$ and outside the outer horizon, in the black hole exterior, if there $\hu+\hw>1/2\,\ln[(1+p)/(1-p)]$. Thus, if $\hu+\hw\leqslant 0$, that is if the sources of the distortion fields satisfy the strong energy condition, then the space-time singularities are located behind the inner (Cauchy) horizon, i.e., there $\eta<-1$ for any $p\in(0,1)$.  

In addition, for a regular horizon there should be no conical singularities on the "semi-axes" $\theta=0$ and $\theta=\pi$, and thus on the horizon. The metric \eq{4.13}--\eq{4.15} has no conical singularities on the "semi-axes" if the space there is locally flat. In other words, the ratio of the Killing vector $\xi^\alpha_{(\phi)}=\delta^\alpha_\phi$ ($\xi^\alpha_{(\chi)}=\delta^\alpha_\chi$) orbit circumference to the orbit radius at the vicinity of the "semi-axis" $\theta=0$ ($\theta=\pi$) should be equal to $2\pi$. For the metric \eq{4.13}--\eq{4.15} this condition implies (for details see \cite{AS-S})
\be\n{4.6}
\hv+2\hu+\hw|_{\theta=0}=0\,,
\ee 
for the "semi-axis" $\theta=0$, and 
\be\n{4.7}
\hv+\hu+2\hw|_{\theta=\pi}=0\,,
\ee
for the "semi-axis" $\theta=\pi$.

Using the expressions \eq{4.2}--\eq{4.5} and the symmetry property of the Legendre polynomials, 
\be\n{4.8}
P_n(-x)=(-1)^nP_n(x)\hh P_n(1)=1\,,
\ee
from the expressions \eq{4.6} and \eq{4.7} we derive the following condition: 
\be\n{4.9}
\sum_{n\geq0}(a_{2n}-b_{2n})+3\sum_{n\geq0}(a_{2n+1}+b_{2n+1})=0\,.
\ee
This condition implies the black hole equilibrium condition.\footnote{Note that different constant of integration, which defines the distortion field $\hv$, would give us exactly the same expression for $\hv$ and the condition \eq{4.9}, after one imposes the no conical singularity conditions \eq{4.6} and \eq{4.7}.}

\section{Properties of the distorted black hole solution}

In this section we shall discuss the distorted black hole solution \eq{4.13}--\eq{4.16} focusing on its horizons. 

\subsection{The inner and outer horizons and their duality transformation}

As it was mentioned in the Introduction, there exists a certain duality transformation between the outer and inner horizon surfaces of a distorted, four-dimensional, static, electrically charged, axisymmetric black hole which is exactly the same as the duality transformation between the horizon and the {\em stretched} singularity surfaces of a four-dimensional, static, vacuum, axisymmetric black hole. Here we show that an analogous situation takes place in a five-dimensional case. We first present the metrics on the three-dimensional surfaces of the outer and inner horizon of the distorted five-dimensional charged black hole and then show the duality transformation between them. This duality transformation is exactly the same as that between the horizon and the {\em stretched} singularity surfaces of a distorted five-dimensional Schwarzschild-Tangherlini black hole.  
 
To begin with, let us introduce the following notations:
\ba
&&u_0=\sum_{n\geq0}a_{2n}\hh u_1=\sum_{n\geq0}a_{2n+1}\hh
w_0=\sum_{n\geq0}b_{2n}\hh w_1=-\sum_{n\geq0}b_{2n+1}\,,\n{5.1}\\
&&u_{\pm}(\theta)=\sum_{n\geq0}(\pm 1)^na_{n}\cos^n\theta-u_0\hh
w_{\pm}(\theta)=\sum_{n\geq0}(\pm 1)^nb_{n}\cos^n\theta-w_0\,.\n{5.2}
\ea
Then, the black hole equilibrium condition \eq{4.9} can be written as
\be\n{5.3}
u_0+3u_1=w_0+3w_1\,,
\ee 
and the strong energy condition \eq{4.12}
takes the form
\be\n{5.4}
u_0+w_0\pm(u_1-w_1)\leqslant 0\,,
\ee
where the sign `$+$' stands for the $\theta=0$ "semi-axis" and the sing `$-$' stands for the $\theta=\pi$ "semi-axis". This condition can be written in the following form:
\be\n{5.4a} 
u_0+w_0\leqslant 0\hh |u_0+w_0|\geqslant
|u_1-w_1|\,.
\ee

Using these notations and the symmetry property of the Legendre polynomials \eq{4.8} we calculate the distortion fields \eq{4.2}--\eq{4.4} at the black hole horizons, 
\ba
&&\hu_{\pm}=u_{\pm}(\theta)+u_{0}\hh
\hw_{\pm}=w_{\pm}(\theta)+w_{0}\,,
\n{5.5}\\
&&\hv_{+}=-\frac{3}{2}(u_{0}+w_{0})-\frac{1}{2}(u_{1}+w_{1})\,,\n{5.6}\\
&&\hv_{-}=-3u_{-}(\theta)-3w_{-}(\theta)-\frac{3}{2}(u_{0}+w_{0})+\frac{1}{2}(u_{1}+w_{1})\,.\n{5.7}
\ea 
 
The three-dimensional surface of the outer horizon is defined by $t=const$ and $\eta=1$. The corresponding metric derived from \eq{4.13}--\eq{4.15} reads
\ba
&&\hspace{2.5cm}d\Sigma^2_{+}
=m(1+p)e^{-(u_0+w_0+3u_1+3w_1)}d\sigma^2_+
\,,\n{5.8}\\
&&\hspace{-1.0cm}d\sigma^2_+=
\frac{1}{4}
\left(e^{2(\cv_++\cu_++\cw_+)}d\theta^2
+2(1+\cos\theta)e^{-2\cw_+}d\chi^2
+2(1-\cos\theta)e^{-2\cu_+}d\phi^2\right)
\,,\n{5.9}
\ea
with the following notations:
\be\n{5.10}
\cu_\pm=u_\pm(\theta)-3u_1\hh
\cw_\pm=w_\pm(\theta)-3w_1\hh
\cv_\pm=4(u_1+w_1)\,.
\ee 
The metric $d\sigma^2_+$ coincides with the metric on the distorted five-dimensional Schwarzschild-Tangherlini black hole horizon surface (see Eq.(96), \cite{AS-S}).

The three-dimensional surface of the inner horizon is defined by $t=const$ and $\eta=-1$. The corresponding dimensionless metric derived from \eq{4.13}--\eq{4.15} reads
\ba
&&\hspace{2.5cm}d\Sigma^2_{-}=m(1-p)e^{(u_0+w_0+3u_1+3w_1)}d\sigma^2_-
\,,\n{5.12}\\
&&\hspace{-1.0cm}d\sigma^2_-
=\frac{1}{4}
\left(e^{-2(\cv_-+\cu_-+\cw_-)}d\theta^2
+2(1+\cos\theta)e^{2\cu_-}d\chi^2
+2(1-\cos\theta)e^{2\cw_-}d\phi^2\right)\,.\n{5.13}
\ea
The metric $d\sigma^2_-$ coincides with the metric on the distorted Schwarzschild-Tangherlini black hole {\em stretched} singularity surface (see Eq.(131), \cite{AS-S}).

The metrics $d\Sigma^2_{+}$ and $d\Sigma^2_{-}$ are related to each other by the following transformation:
\ba
&&u_0\longrightarrow -w_0\hh
u_1\longrightarrow -w_1\hh
u_\pm\longrightarrow -w_\mp\,,\n{5.15}\\
&&w_0\longrightarrow -u_0\hh
w_1\longrightarrow -u_1\hh
w_\pm\longrightarrow -u_\mp\,,\n{5.16}\\
&&\hspace{2.8cm}\pm p\longrightarrow \mp p
\,.\n{5.16a}
\ea
Note that according to the expression \eq{2.12a}, the transformation \eq{5.16a} corresponds to a "switch" between the outer and inner horizons of the Reissner-Nordstr\"om black hole. The no conical singularity condition \eq{5.3} does not change under the transformation \eq{5.15}--\eq{5.16}. However, this transformation implies a "switch" between the "semi-axes" $\theta=0$ and $\theta=\pi$ and the reverse of the inequality sign in the strong energy condition \eq{5.4}. This is because Killing vector $\xi^\mu_{(t)}$ is space-like in the region between the horizons, that implies negative energy. The transformation \eq{5.15}--\eq{5.16} implies the following transformation between the multipole moments:
\be\n{5.17}
a_{2n} \longrightarrow -b_{2n}\hh 
a_{2n+1} \longrightarrow b_{2n+1}\hh 
b_{2n} \longrightarrow -a_{2n}\hh 
b_{2n+1} \longrightarrow a_{2n+1} \,.
\ee
Such a transformation corresponds to an exchange between the "semi-axes" and reverse of signs of the multipole moments,
\be\n{5.18}
(\theta,\chi,\phi)\longrightarrow 
(\pi-\theta,\phi,\chi)\hh a_n\longrightarrow-a_n\hh b_n\longrightarrow-b_n\,.
\ee
We shall call the transformation \eq{5.15}--\eq{5.18} the {\em duality transformation} between the outer and inner horizons of the distorted black hole. The duality transformation \eq{5.17}--\eq{5.18} is exactly the same as the duality transformation between the horizon and the {\em stretched} singularity surfaces of the distorted five-dimensional Schwarzschild-Tangherlini black hole (see Eqs. (135)--(137), \cite{AS-S}). 

\subsection{Surface gravity, electrostatic potential, horizon areas product and inequality, the Smarr formula}

The surface gravity at the horizons is
\be\n{5.22}
\kappa_\pm=\frac{2p\,e^{\pm \frac{3}{2}\gamma}}{\sqrt{m(1\pm p)^3}}\,,
\ee
where
\be\n{5.20}
\gamma=u_0+w_0+\frac{1}{3}(u_1+w_1)\,.
\ee
According to the strong energy condition \eq{5.4a}, we have $\gamma\leqslant 0$. We see that due to distortion fields, the surface gravity differs from that of the Reissner-Nordstr\"om (undistorted) black hole \eq{2.16} by the factor $e^{\pm \frac{3}{2}\gamma}$. One can interpret this factor as an "effective redshift factor" due to the distortion fields. However, we should bear in mind that the distortion fields contribute to both the local acceleration (acceleration of the Killing field $\BM{\xi}_{(t)}$ orbit) and to the redshift factor of the space-time, $\sqrt{-g_{tt}}\,$.

The horizon areas of the distorted black hole solution \eq{4.13}--\eq{4.16} are 
[cf. Eq.\eq{2.15}]
\be\n{5.19}
\mathcal{A}_\pm=2\pi^2
\sqrt{m^3(1\pm p)^3}\,e^{\mp \frac{3}{2}\gamma}\,.
\ee
One can see that the area product 
\be\n{5.21}
\mathcal{A}_+\mathcal{A}_-=4\pi^4q^3
=\frac{\pi^4}{6\sqrt{3}}Q^3\,.
\ee
has the same form as that of the Reissner-Nordstr\"om black hole \eq{2.15a}. Let us define the lower and upper limits for the values of $\mathcal{A}_{+}$ and $\mathcal{A}_{-}$, respectively. In order to do so, we calculate the ratio $Q^{3/2}/\mathcal{A}_{\pm}$. Using the expressions \eq{2.8}, \eq{2.12}, and \eq{5.19} we derive
\be\n{ain}
\frac{Q^{\frac{3}{2}}}{\mathcal{A}_{-}}=\frac{3^{\frac{3}{4}}\sqrt{2}}{\pi^2}\left(\frac{1+p}{1-p}\right)^{\frac{3}{4}}e^{-\frac{3}{2}\gamma}\hh \frac{Q^{\frac{3}{2}}}{\mathcal{A}_{+}}=\frac{3^{\frac{3}{4}}\sqrt{2}}{\pi^2}\left(\frac{1-p}{1+p}\right)^{\frac{3}{4}}e^{\frac{3}{2}\gamma}\,.
\ee
Then, the conditions $\gamma\leqslant 0$ and $p\in(0,1)$ allow us to define the lower and upper limits for the inner and outer horizon areas of a general distorted Reissner-Nordstr\"om black hole,
\be\n{ain2}
\mathcal{A}_{-}<\frac{\pi^2 Q^\frac{3}{2}}{3^{\frac{3}{4}}\sqrt{2}}<\mathcal{A}_{+}\,,
\ee
Using the area product \eq{5.21} we present the universal area inequality \eq{ain2} in the following form:
\be\n{5.21a}
\mathcal{A}_{-}<\sqrt{\mathcal{A}_{-}
\mathcal{A}_{+}}<\mathcal{A}_{+}\,.
\ee
Thus, the geometric mean of the inner and outer horizon areas of the distorted black hole represents the upper and lower limits of its inner and outer horizon areas, respectively.\footnote{Hennig, Cederbaum, and Ansorg \cite{Hennig1} have proven a similar inequality for the areas of regular axisymmetric and stationary four-dimensional sub-extremal black hole with surrounding matter in the Einstein-Maxwell theory.}

The electrostatic potential \eq{4.16} takes the following values at the black hole horizons [cf. Eq.\eq{2.16a}]
\be\n{5.22b}
\Phi_\pm=\frac{\sqrt{3(1-p^2)}}{1\pm p}\,.
\ee
Thus, the values of the electrostatic potential at the black hole horizons does not change under the distortion. 
Calculating the electromagnetic field invariant $F^2$ at the black hole horizons we find
\be\n{5.23}
F^2_\pm=-\frac{24(1-p^2)}{m(1\pm p)^3}\,
e^{\pm3\gamma}\,.
\ee
Using this expression and the expression \eq{5.22} one can show that the relation [cf. Eq.\eq{2.18}] 
\be\n{5.23a}
F^2_\pm=-\frac{6}{p^2}(1-p^2)\kappa^2_\pm\,.
\ee
holds for the distorted black hole as well. Note that the quantities \eq{5.19}, \eq{5.22}, \eq{5.22b}, and \eq{5.23} are related by the duality transformation \eq{5.15}--\eq{5.16a}.

One can check that the Smarr formula [cf. Eq.\eq{2.17}]
\be\n{5.22a}
\pm M=\frac{3}{16\pi}\kappa_\pm \mathcal{A}_\pm\pm\frac{\pi}{8}\Phi_\pm Q\,.
\ee
holds for the distorted black hole as well. Here $M$ defines the black hole Komar mass \eq{2.9}, assuming that the space-time \eq{4.13}--\eq{4.15} can be analytically extended to achieve its asymptotic flatness.

\section{Mechanics and thermodynamics of the distorted black hole}

Mechanical laws of black holes represent relations between the black hole variables, such as mass, horizon area, surface gravity, etc  (see \cite{BCH}). These variables, in turn, correspond to the thermodynamic variables, such as energy, entropy, temperature, etc. This correspondence is due to Hawking radiation \cite{Hawk}, which "endows" a black hole with temperature.\footnote{Such a correspondence had previously been conjectured by Bekenstein \cite{Bek}.} In this section, we derive mechanical laws of the distorted black hole and present the corresponding laws of thermodynamics. We shall mostly follow the arguments of Geroch and Hartle \cite{Ger.IV}, who constructed the global and local forms of the first law for a four-dimensional distorted vacuum black hole. The results of Geroch and Hartle were generalized by Fairhurst and Krishnan \cite{Fai.IV} for an electrically charged four-dimensional black hole distorted by external charged matter.

\subsection{The zeroth law}

The zeroth law says that a black hole surface gravity (and accordingly, its temperature) is constant at the black hole horizon. The surface gravity is defined up to an arbitrary constant which depends on the normalization of the time-like Killing vector. However, the normalization does not affect the zeroth law. As we readily see from the expression \eq{5.22}, the zeroth law holds for both the horizons of our distorted black hole. The corresponding temperature is defined in terms of the surface gravity as 
\be\n{t1}
T_{\pm}=\frac{\kappa_{\pm}}{2\pi}\,.
\ee
This definition, however, requires a proper normalization of the Killing vector at the spatial infinity. 

Though the temperature $T_{+}$ is associated with the black hole outer horizon is a typical quantity, the "temperature" $T_{-}$, which is associated with the black hole inner (Cauchy) horizon, is rather a dubious thermodynamic variable. However, taking into account the description of black holes within string theory \cite{Hor}, one can view the outer horizon thermodynamics as the sum of the thermodynamics corresponding to the left- and right-moving excitations of the string. Due to such a duplicate nature of the horizon thermodynamics, one can view the inner horizon thermodynamics as the difference of the thermodynamics corresponding to the right- and left-moving excitations of the string \cite{Lar,CvetLar,CvetLar2}. In this picture, the thermodynamic variables corresponding to the outer horizon are mapped to the thermodynamic variables corresponding to the inner horizon. In our case, such a map is represented by the duality transformation (see Eqs.\eq{5.15}--\eq{5.16a} and the comment at the very end of the previous Section). We will follow this picture in our description of thermodynamics of the distorted black hole. 

The electrostatic potential \eq{5.22b} is constant at the black hole horizon. Note that this condition holds for more general class of regular horizons (see, e.g., \cite{ASEMdL}).

\subsection{The first law}

The first law of the black hole mechanics represents a relation between the black hole's two nearby equilibrium configurations related by a change in the black hole mass, horizon area, and other black hole parameters, e.g., electric charge and angular momentum, as well as the change in the stress-energy of the external matter, if present. There are different forms of the first law, which are defined according to the system in question. Here we shall consider global and local forms of the first law which correspond to the total system of the black hole plus the distorting matter and to the black hole alone, respectively.

\subsubsection{The global first law}

The global first law is defined for the system which consists of the black hole and the external matter acting on it. To define the global first law of black hole mechanics, we need to extend the space-time \eq{4.13}-\eq{4.15} to achieve its asymptotic flatness. Such an extension is possible assuming that one can include into the solution the sources of the distorting matter. Accordingly, the Einstein-Maxwell equations get violated in the region of the sources location due to their energy-momentum tensor. The extension is achieved by requiring that the distortion fields $\hu,\hw$, and $\hv$ vanish  at the asymptotic infinity and by extending the corresponding space-time manifold. In the extended manifold there exists an electrovacuum region in the interior of the black hole and part of the exterior region where the solution \eq{4.13}--\eq{4.16} is valid. Then, there is a region where the  external sources are located. Beyond that region there is asymptotically flat electrovacuum region. Having this extension one can normalize the timelike Killing vector $\BM{\xi}_{(t)}$ at the spatial infinity as $\BM{\xi}_{(t)}^2=-1$. As it is done, one  naturally finds the Komar mass $M$ [see Eqs. \eq{2.8}--\eq{2.9}] of the distorted black hole equal to the Komar mass of the Reissner-Nordstr\"om (undistorted) black hole. Then, using the expression for the black hole horizon area \eq{5.19} one can find the relation $M=M(\mathcal{A}_\pm,Q,\gamma)$. Differentiating $M$ with respect to its arguments and using the expressions for the surface gravity \eq{5.22}, the electrostatic potential \eq{5.22b}, and for the black hole local mass, which can be calculated by using the definition \eq{2.9} with the three-dimensional closed space-like surface $S_{(3)}$ at the black hole horizons, we derive the global first law of the black hole mechanics,
\be\n{t2}
\pm\delta M=\frac{\kappa_\pm}{8\pi}\delta \mathcal{A}_\pm
\pm\frac{\pi}{8}\Phi_\pm\delta Q+M_\pm^{\text{loc}}\delta\gamma\,,
\ee
where the local black hole mass,
\be\n{t3}
M_\pm^{\text{loc}}=\pm\frac{3\pi}{4}mp\,,
\ee
does not depend on the distortion fields. 

From the global first law of the black hole mechanics, by using the definition of temperature \eq{t1} and the black hole entropy,
\be\n{t4}
S_\pm=\frac{\mathcal{A}_\pm}{4}\,,
\ee
we derive the global first law of the black hole thermodynamics,
\be\n{t5}
\pm\delta M=T_\pm\delta S_\pm
\pm\frac{\pi}{8}\Phi_\pm\delta Q+M_\pm^{\text{loc}}\delta\gamma\,.
\ee       
Here the term $M_\pm^{\text{loc}}\delta\gamma$ is interpreted as the work done on the black hole by the variation of the external potential $\gamma$ due to the distorting matter. If the distortion is adiabatic, $\delta S_\pm=0$, i.e., such that neither matter nor gravitational waves cross the black hole horizons, and in addition, the black hole charge $Q$ does not change, then the work $M_\pm^{\text{loc}}\delta\gamma$ results in the change of the black hole mass $\delta M$. 
  
\subsubsection{The local first law}

The local first law is defined for the system which consists of the black hole alone. The local first law does not include the distorting matter into consideration of the black hole mechanics. It can be defined by observers who live near the black hole and attribute the local gravitational field to the black hole alone. These observers consider the black hole as an isolated, undistorted object.\footnote{Note that according to the expressions \eq{5.19}, \eq{5.22}, \eq{5.22b}, and \eq{5.23}, all deviations from spherical symmetry vanish at the black hole horizons.} Thus, assuming that there is no other matter present and the space-time is asymptotically flat, they define its surface gravity $\tilde{\kappa}_+$, the outer horizon area $\mathcal{\tilde A}_+$, electrostatic potential $\tilde{\Phi}_+$, electric charge $\tilde{Q}$, and the black hole Komar mass $\tilde{M}$ such that they satisfy the Smarr formula for the Reissner-Nordstr\"om black hole,
\be\n{t6a}
\tilde{M}=\frac{3}{16\pi}\tilde{\kappa}_+ \mathcal{\tilde{A}}_++\frac{\pi}{8}\tilde{\Phi}_+ \tilde{Q}\,.
\ee
Accordingly, the Smarr formula for the black hole Cauchy horizon reads [cf. Eq.\eq{2.17}]
\be\n{t6b}
-\tilde{M}=\frac{3}{16\pi}\tilde{\kappa}_- \mathcal{\tilde{A}}_--\frac{\pi}{8}\tilde{\Phi}_- \tilde{Q}\,.
\ee

Note, that because the distortion fields do not carry electrostatic energy, we have $\tilde{\Phi}_\pm=\Phi_\pm$ [see Eq.\eq{5.22b}] and $\tilde{Q}=Q$. Thus, these observers construct the local first law of the black hole mechanics as that of the Reissner-Nordstr\"om (undistorted) black hole, 
\be\n{t7}
\pm\delta \tilde{M}=\frac{\tilde{\kappa}_\pm}{8\pi}\delta \mathcal{\tilde{A}}_\pm
\pm\frac{\pi}{8}\Phi_\pm\delta Q\,.
\ee
With the definitions of temperature \eq{t1} and entropy \eq{t4} the local first law of black hole thermodynamics reads
\be\n{t8}
\pm\delta \tilde{M}=\tilde{T}_\pm\delta \tilde{S}_\pm
\pm\frac{\pi}{8}\Phi_\pm\delta Q\,.
\ee 

The measurements of the observers define the black hole area as that which is exactly equal to the black hole area when the presence of the distortion fields is taken into account, i.e.,
\be\n{t9}
\mathcal{\tilde{A}}_\pm=2\pi^2
\sqrt{\tilde{m}^3(1\pm \tilde{p})^3}=\mathcal{A}_{\pm}=2\pi^2
\sqrt{m^3(1\pm p)^3}\,e^{\mp \frac{3}{2}\gamma}\,,
\ee
where [cf. Eqs.\eq{2.8} and \eq{2.12}]
\be\n{t10}
\tilde{M}=\frac{3\pi}{4}\tilde{m}\hh \tilde{p}=\frac{1}{\tilde{m}}\sqrt{\tilde{m}^2-q^2}\,.
\ee
The expression \eq{t9} gives us the following relations:
\ba
\tilde{M}&=&\frac{M}{2}\left[(1+p)e^{-\gamma}+(1-p)e^{\gamma}\right]\,,\n{t11}\\
\tilde{p}&=&\frac{(1+p)e^{-\gamma}-(1-p)e^{\gamma}}{(1+p)e^{-\gamma}+(1-p)e^{\gamma}}\,.\n{t12}
\ea
Using these relations and the expressions for the surface gravity \eq{2.16}, \eq{5.22} and for the electrostatic potential \eq{2.16a}, \eq{5.22b} we derive
\ba
\tilde{\kappa}_{\pm}&=&\frac{2\tilde{p}}{\sqrt{\tilde{m}(1\pm\tilde{p})^{3}}}=\frac{\kappa_{\pm}}{2p}\left[(1+p)e^{-\gamma}-(1-p)e^{\gamma}\right]\,,\n{t13}\\
\tilde{\Phi}_{\pm}&=&\frac{\sqrt{3(1-\tilde{p}^2)}}{1\pm \tilde{p}}=\Phi_{\pm}e^{\gamma}\,,\n{t14}
\ea
where $\kappa_{\pm}$ and $\Phi_{\pm}$ are given by the expressions \eq{5.22} and \eq{5.22b}, respectively. The relations \eq{t11}--\eq{t14} provide us with the correspondence between the local and the global forms of the first law. Namely, substituting these relations into the expressions \eq{t7} and \eq{t8} we derive the expressions \eq{t2} and \eq{t5}.

\section{Space-time curvature at the horizons}

Let us now calculate space-time curvature at the distorted black hole horizons. The main purpose of such a calculation is to analyse the effect of the distortion fields on the inner horizon. Here we shall calculate the Kretschmann scalar (the trace of the square of the Riemann tensor) value at the outer and inner horizons. As an example, we consider the dipole-monopole and quadrupole-quadrupole distortion fields. Using the results of our previous work \cite{ASEMdL} we can write the Kretschmann scalar at the horizons as follows (see Eq.(65) of \cite{ASEMdL} for $d=5$)\footnote{Note that the Ricci scalar and the trace of the square of the Ricci tensor, $R_{\alpha\beta}R^{\alpha\beta}$, calculated at the horizons are proportional to $F^2$ value at the horizons (see Eqs.(63)--(64) of \cite{ASEMdL}, setting the cosmological constant and dilaton field to zero). Thus, if they diverge, then the Kretschmann scalar diverges too.}: 

\be\n{5.24}
\mathcal{K}_\pm
=6(\mathcal{R}_{AB}\mathcal{R}^{AB})_{\pm}
+\frac{3}{2}F^2_\pm\mathcal{R}_\pm
+\frac{55}{144}F^4_\pm\,,
\ee
where $(\mathcal{R}_{AB}\mathcal{R}^{AB})_{\pm}$ is the trace of the square of the Ricci tensor of the horizon surfaces,
$\mathcal{R}_{\pm}$ is the Ricci scalar of the horizon surfaces, and $F_{\pm}$ is the electromagnetic field invariant calculated at the horizons. Here and in what follows, we shall use the capital Latin letters for three-dimensional objects defined on the horizon surfaces. Using the relation \eq{5.23a} we derive
\be\n{5.25}
\mathcal{K}_\pm
=6(\mathcal{R}_{AB}\mathcal{R}^{AB})_{\pm}
-\frac{9}{p^2}(1-p^2)\kappa^2_\pm\mathcal{R}_\pm
+\frac{55}{4p^4}(1-p^2)^2\kappa^4_\pm\,.
\ee
Thus, for an arbitrary static and "axisymmetric" distortion, the value of the Kretschmann scalar at the horizons is defined by the horizons intrinsic curvature and surface gravity. The corresponding Ricci scalar and the trace of the square of the Ricci tensor can be expressed through the Ricci tensor components as follows:
\be\n{5.26}
\mathcal{R}_{\pm}=\mathcal{R}^{\,\phi}_{\pm\phi}+\mathcal{R}^{\,\chi}_{\pm\chi}+\mathcal{R}^{\,\theta}_{\pm\theta}\hh
(\mathcal{R}_{AB}\mathcal{R}^{AB})_{\pm}
=(\mathcal{R}^{\,\phi}_{\pm\phi})^2
+(\mathcal{R}^{\,\chi}_{\pm\chi})^{2}
+(\mathcal{R}^{\,\theta}_{\pm\theta})^2\,.
\ee
Components of the Ricci tensor corresponding to a three-dimensional surface, in turn, can be expressed through the Gaussian curvatures $K_{\theta}$, $K_{\chi}$, and $K_{\phi}$ of the two-dimensional sections $(\chi,\phi)$, $(\theta,\phi)$, and $(\theta,\chi)$ as (see, e.g., \cite{Eisen,FR})
\be\n{5.27} 
\mathcal{R}^{\phi}_{\pm\phi}
=K_{\pm\theta}+K_{\pm\chi}\hh
\mathcal{R}^{\chi}_{\pm\chi}
=K_{\pm\theta}+K_{\pm\phi}\hh
\mathcal{R}^{\theta}_{\pm\theta}
=K_{\pm\chi}+K_{\pm\phi}\,.
\ee
The Gaussian curvatures of the sections are expressed through the corresponding Riemann tensor components of the metrics $d\Sigma^2_+$, Eq.\eq{5.8}, and $d\Sigma^2_-$, Eq.\eq{5.12}, calculated in an orthonormal frame $x^{\hat{\alpha}}$ (see, e.g., \cite{Eisen,FR}), 
\ba\n{5.28}
&&\hspace{-1.2cm}K_{+\phi}=\mathcal {R}_{+\hat{\theta}\hat{\chi}\hat{\theta}
\hat{\chi}}=\mathcal{N}_+
\biggl[1+4w_{+,\theta\theta}-8w_{+,\theta}^2-4u_{+,\theta}w_{+,\theta}
-\frac{2\sin\theta}{1+\cos\theta}(u_{+,\theta}+3w_{+,\theta})\biggr]\,,
\non\\
&&\hspace{-1.2cm}K_{+\chi}=\mathcal {R}_{+\hat{\theta}\hat{\phi}\hat{\theta}
\hat{\phi}}=\mathcal{N}_+
\biggl[1+4u_{+,\theta\theta}-8u_{+,\theta}^2-4u_{+,\theta}w_{+,\theta}
+\frac{2\sin\theta}{1-\cos\theta}(w_{+,\theta}+3u_{+,\theta})\biggr]\,,
\\
&&\hspace{-1.2cm}K_{+\theta}=\mathcal {R}_{+\hat{\chi}\hat{\phi}\hat{\chi}
\hat{\phi}}=\mathcal
{N}_+\biggl[1-4u_{+,\theta}w_{+,\theta}-\frac{2}{\sin\theta}(u_{+,\theta}-w_{+,\theta})
+2\cot\theta(u_{+,\theta}+w_{+,\theta})\biggr]\,,
\non
\ea
for the outer horizon surface, and
\ba\n{5.29}
&&\hspace{-1.2cm}K_{-\phi}=\mathcal {R}_{-\hat{\theta}\hat{\chi}\hat{\theta}
\hat{\chi}}=\mathcal{N}_-
\biggl[1-4u_{-,\theta\theta}-8u_{-,\theta}^2-4u_{-,\theta}w_{-,\theta}
+\frac{2\sin\theta}{1+\cos\theta}(w_{-,\theta}+3u_{-,\theta})\biggr]\,,
\non\\
&&\hspace{-1.2cm}K_{-\chi}=\mathcal {R}_{-\hat{\theta}\hat{\phi}\hat{\theta}
\hat{\phi}}=\mathcal{N}_-
\biggl[1-4w_{-,\theta\theta}-8w_{-,\theta}^2-4u_{-,\theta}w_{-,\theta}
-\frac{2\sin\theta}{1-\cos\theta}(u_{-,\theta}+3w_{-,\theta})\biggr]\,,
\\
&&\hspace{-1.2cm}K_{-\theta}=\mathcal {R}_{-\hat{\chi}\hat{\phi}\hat{\chi}
\hat{\phi}}=\mathcal
{N}_-\biggl[1-4u_{-,\theta}w_{-,\theta}+\frac{2}{\sin\theta}(w_{-,\theta}-u_{-,\theta})
-2\cot\theta(w_{-,\theta}+u_{-,\theta})\biggr]\,,
\non
\ea
for the inner horizon surface. Here 
\be\n{5.30}
\mathcal{N}_\pm=\frac{1}{m(1\pm p)}e^{\mp 2[u_\pm(\theta)+w_\pm(\theta)]\pm(u_1+w_1+
u_0+w_0)}\,.
\ee
Thus, defining the values of the distortion fields $\hu,\hw$, and $\hv$ at the black hole horizons one can calculate the Gaussian curvatures \eq{5.28}--\eq{5.29} and using the expressions \eq{5.26}--\eq{5.27} find ${\cal R}_\pm$ and  $({\cal R}_{AB}{\cal R}^{AB})_\pm$. Note that in the case of the distortion fields $\hu=0$, $\hw\ne0$ we have $K_{+\chi}=K_{+\theta}$, $K_{-\phi}=K_{-\theta}$, and in the case of the distortion fields $\hu\ne0$, $\hw=0$ we have $K_{+\phi}=K_{+\theta}$, $K_{\chi}=K_{-\theta}$. For a round three-dimensional sphere, which represents the horizon surfaces of a five-dimensional Reissner-Nordstr\"om black hole \eq{2.13}--\eq{2.14} we have
\be\n{5.31}
K_{\pm\phi}=K_{\pm\chi}=K_{\pm\theta}
=\frac{1}{m(1\pm p)}\,.
\ee

We see from the expression \eq{5.25} that if the distortion fields are regular and smooth at the outer horizon, then the space-time curvature at the outer horizon is regular. We can see that the expressions \eq{5.28} and \eq{5.29} are related by the duality transformation \eq{5.15}--\eq{5.16a}, and therefore, the quantities $(\mathcal{R}_{AB}\mathcal{R}^{AB})_{\pm}$ and $\mathcal{R}_{\pm}$. As we mentioned at the end of Section 5, the expressions for the surface gravity calculated at the horizons, \eq{5.22}, are related by the duality transformation as well. Thus, according to the expression \eq{5.25}, the Kretschmann scalars calculated at the outer and inner horizons are related by the duality transformation. This implies that if $\mathcal{K}_+$ is regular, then $\mathcal{K}_-$ is regular as well.

\section{Analysis of the distorted black hole}

\subsection{The model}

After the study of the properties of the charged distorted black hole solution, let us now discuss our model. We shall consider a five-dimensional Reissner-Nordstr\"om black hole which is distorted adiabatically by the external matter. More precisely, according to the solution \eq{4.1}--\eq{4.5}, the external matter is located at the asymptotic infinity. The matter is described by the multipole moments. For the vanishing multipole moments the solution is the Reissner-Nordstr\"om black hole of the given mass and charge defined by the parameters $m_0$ and $p_0$. For a slow (adiabatic) change of the multipole moments, such that neither gravitational waves nor matter cross the black hole outer horizon, the horizon area $\mathcal{A}_+$ does not change. Then, according to the relation \eq{2.15a}, the area of the inner black hole horizon $\mathcal{A}_-$ does not change either. Thus, for arbitrary values of the multipole moments we have [see Eqs.\eq{2.15} and \eq{5.19}]
\ba
&&m(1\pm p)e^{\mp\gamma}=m_0(1\pm p_0)\,,\n{6.1}\\
&&m^2(1-p^2)=q^2=m_0^2(1-p_0^2)\,.\n{6.2}
\ea
Therefore, we can express the factor $\mathcal{N}_\pm$ [see Eq.\eq{5.30}] and the electromagnetic field invariant \eq{5.23} as follows:
\ba
&&\mathcal{N}_\pm=\frac{1}{m_0(1\pm p_0)}e^{\mp 2[u_\pm(\theta)+w_\pm(\theta)]\pm\frac{2}{3}(u_1+w_1)}\,,\n{6.3}\\
&&F^2_\pm=-24\frac{(1-p_0^2)}{m_0(1\pm p_0)^3}\,.\n{6.4}
\ea
Thus, the expression for the Kretschmann scalar \eq{5.24} can be written as
\be\n{6.5}
\mathcal{K}_\pm
=6(\mathcal{R}_{AB}\mathcal{R}^{AB})_{\pm}
-36\frac{(1-p_0^2)}{m_0(1\pm p_0)^3}
\mathcal{R}_\pm
+220\frac{(1-p_0^2)^2}{m_0^2(1\pm p_0)^6}\,,
\ee
where the quantities $(\mathcal{R}_{AB}\mathcal{R}^{AB})_{\pm}$ and $\mathcal{R}_\pm$ are calculated with the use of \eq{5.26}--\eq{5.29} where the factor $\mathcal{N}_\pm$ is given by \eq{6.3}. Note that the expression \eq{6.5} does not depend on the monopole moments $a_0$, $b_0$.

We would like to analyze the effect of the distortion fields on the black hole horizons. For this purpose, we shall calculate ${\cal K}_{\pm}$ in units of ${\cal K}_{\text{RN}\pm}$ corresponding to the Reissner-Nordstr\"om black hole. Using the expressions
\eq{2.16}, \eq{5.25}--\eq{5.27}, and \eq{5.31}, we derive
\be\n{6.6}
\mathcal{K}_{\text{RN}\pm}=\frac{4}{m_0^2(1\pm p_0)^4}
\left(19\mp 74p_0+127p_0^2\right)\,,
\ee 
and construct the following function:
\be\n{6.7}
k_{\pm}=
\frac{{\mathcal{K}_\pm}}{\mathcal{K}_{\text{RN}\pm}}\,.
\ee
This function illustrates the relative space-time curvature at the distorted black hole horizons. In the following subsection, we shall analyze this function on the example of two simple and dominant cases of the dipole-dipole and quadrupole-quadrupole distortion fields.

\subsection{An example: The dipole-monopole and quadrupole-quadrupole distortions}

Let us now illustrate the effect of the distortion fields on the black hole horizons by considering two simple examples of the distortion fields -- the dipole-monopole and the quadrupole-quadrupole distortions, where the first name stands for the $\hu-$field and the second name sands for the $\hw-$field. For the dipole-monopole distortion we have [see Eqs.\eq{5.1}--\eq{5.4a}]
\ba\n{6.8}
&&u_0=a_0\hh u_1=a_1\hh u_{\pm}(\theta)=\pm a_1\cos\theta\hh w_0=a_0+3a_1\hh w_1=0 \hh
w_{\pm}(\theta)=0\,,\non\\
&&\hspace{4.0cm}2a_0+3a_1\leqslant 0\hh |2a_0+3a_1|\geqslant |a_1|\,.
\ea
Substituting these expressions into \eq{5.28}--\eq{5.29} with $\mathcal{N}_\pm$ given by \eq{6.3} we derive
\be\n{6.9}
K_{+\phi}=K_{+\theta}
=k^{\text{dm}}_{1+}\hh
K_{+\chi}=k^{\text{dm}}_{2+}\hh
K_{-\chi}=K_{-\theta}
=k^{\text{dm}}_{1-}\hh
K_{-\phi}=k^{\text{dm}}_{2-}\,,
\ee
where
\ba\n{6.10}
k^{\text{dm}}_{1\pm}&=&
\frac{e^{-2a_1\cos\theta\pm\frac{2}{3}a_1}}{m_0(1\pm p_0)}
\left(1\pm 2a_1-2a_1\cos\theta\right)\,,\non\\
k^{\text{dm}}_{2\pm}&=&
\frac{e^{-2a_1\cos\theta\pm\frac{2}{3}a_1}}{m_0(1\pm p_0)}
\left(1\mp 6a_1-10a_1\cos\theta
-8a_1^2\sin^2\theta\right)\,.
\ea
For the quadrupole-quadrupole distortion we have [see Eqs.\eq{5.1}--\eq{5.4a}]
\be\n{6.11}
u_{0}=w_{0}=a_{0}+a_{2}\hh u_{1}=w_{1}=0\hh u_{\pm}(\theta)=w_{\pm}(\theta)=-a_{2} \sin^{2} \theta\hh
a_0+a_2\leqslant 0\,.
\ee
Substituting these expressions into \eq{5.28}--\eq{5.29} with $\mathcal{N}_\pm$ given by \eq{6.3} we derive
\ba\n{6.12}
K_{\pm\phi}&=&k^{\text{qq}}_{1\pm}\hh K_{\pm\chi}=k^{\text{qq}}_{2\pm}\hh
K_{\pm\theta}=k^{\text{qq}}_{3\pm}\,,
\ea
where
\ba\n{6.13}
k^{\text{qq}}_{1\pm}&=&
\frac{e^{\pm 4a_2\sin^2\theta}}
{m_0(1\pm p_0)}
\left(1\pm 8a_{2}(1+2\cos\theta-4\cos^{2}\theta)
-48a^{2}_{2}\cos^{2}\theta\sin^{2}\theta
\right)\,,\non\\
k^{\text{qq}}_{2\pm}&=&
\frac{e^{\pm 4a_2\sin^2\theta}}
{m_0(1\pm p_0)}
\left(1\pm 8a_{2}(1-2\cos\theta-4\cos^{2}\theta)
-48a^{2}_{2}\cos^{2}\theta\sin^{2}\theta
\right)\,,\non\\
k^{\text{qq}}_{3\pm}&=&
\frac{e^{\pm 4a_2\sin^2\theta}}
{m_0(1\pm p_0)}
\left(1\mp 8a_{2}\cos^{2}\theta
-16a^{2}_{2}\cos^{2}\theta
\sin^{2}\theta\right)\,.
\ea
\begin{figure}[htb]
\begin{center}
\hspace{0cm}
\includegraphics[width=12cm]{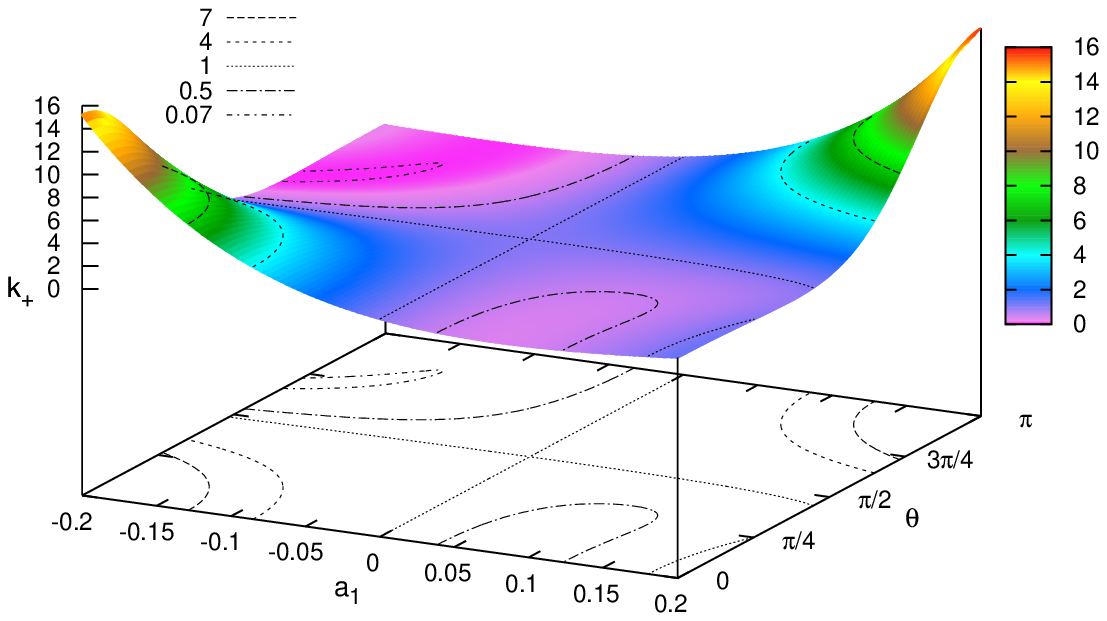}
\caption{The dipole-monopole distortion, $p_0=0.6$. The Kretschmann scalar at the outer horizon of the distorted black hole, in units of that of the undistorted one.}\label{F1}
\end{center}
\end{figure}

\begin{figure}[htb]
\begin{center}
\hspace{0cm}
\includegraphics[width=12cm]{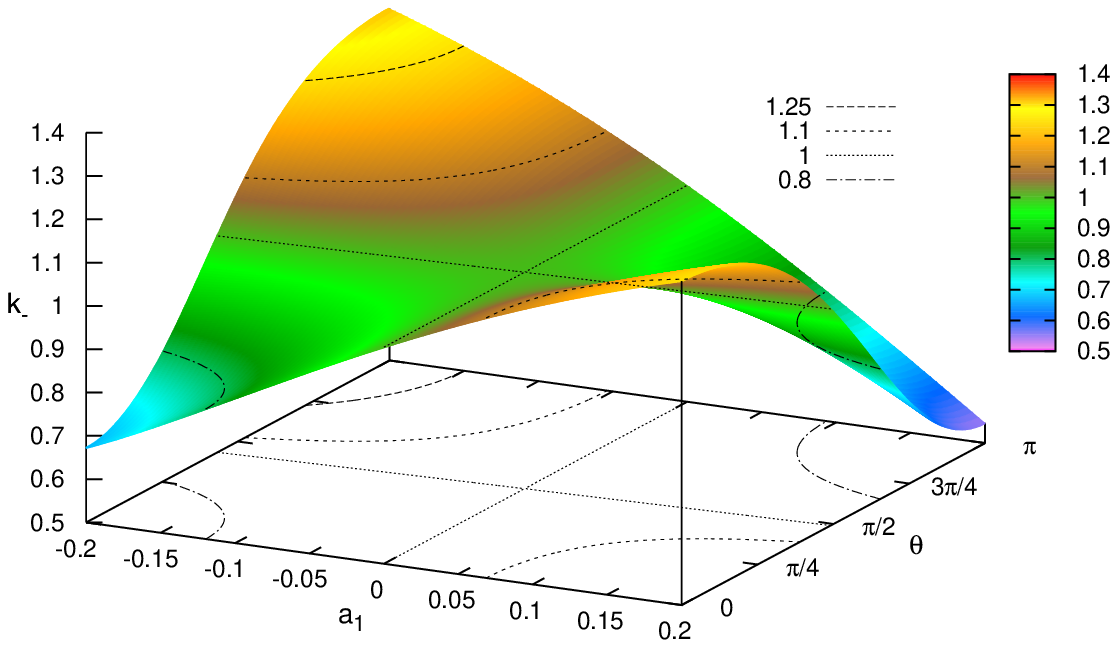}
\caption{The dipole-monopole distortion, $p_0=0.6$. The Kretschmann scalar at the inner horizon of the distorted black hole, in units of that of the undistorted one.}\label{F2}
\end{center}
\end{figure}

\begin{figure}[htb]
\begin{center}
\hspace{0cm}
\includegraphics[width=12cm]{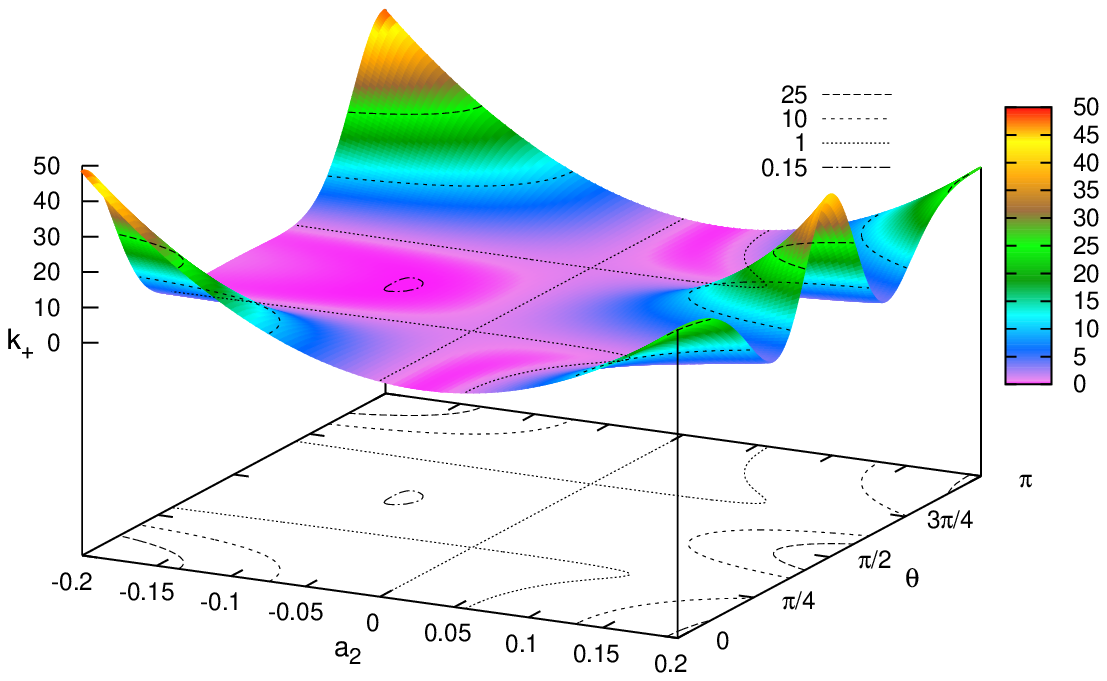}
\caption{The quadrupole-quadrupole distortion, $p_0=0.6$. The Kretschmann scalar at the outer horizon of the distorted black hole, in units of that of the undistorted one.}\label{F3}
\end{center}
\end{figure}

\begin{figure}[htb]
\begin{center}
\hspace{0cm}
\includegraphics[width=12cm]{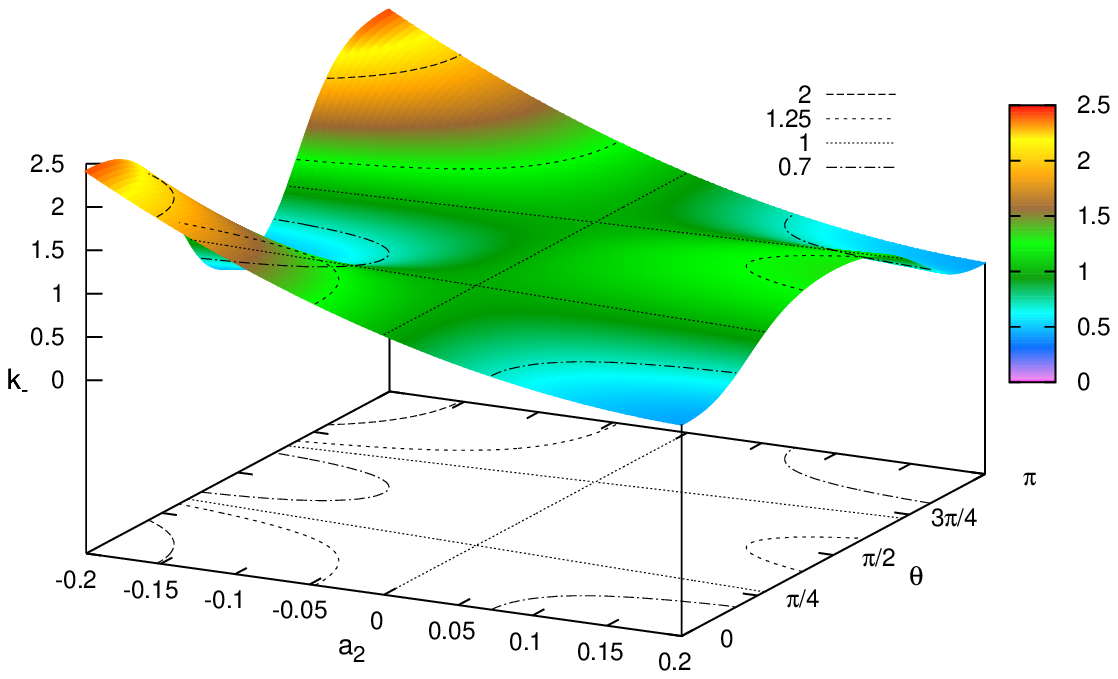}
\caption{The quadrupole-quadrupole distortion, $p_0=0.6$. The Kretschmann scalar at the inner horizon of the distorted black hole, in units of that of the undistorted one.}\label{F4}
\end{center}
\end{figure}
Using these expressions, one can evaluate the function $k_{\pm}$ [see Eq.\eq{6.7}]. This function is plotted in Figs.~\ref{F1}--\ref{F4} for $p_0=0.6$, that corresponds to $q_0/m_0=0.8$. From the plots we see that as the result of the distortion, the Kretschmann scalar varies over the horizon surface, and for the given range of the dipole and quadrupole moments the maximal values of $k_+$ are 10~---~20 times greater than those of $k_-$. This implies that the outer horizon is more susceptible to the distortion. In the case of the quadrupole-quadrupole distortion, for a given value of $p_{0}$ there exits a local minima of $k_{\pm}$ at $\theta=\pi/2$ for some value of $a_{2}$ given by the root of the expression
\be
(14\pm 112a_{2}+192a_{2}^{2})(1\pm p_0)e^{\pm4a_{2}}-(21\pm 48a_{2})(1\mp p_0)=0.
\ee
For example, for $p_{0}=0.6$ the local minimum $k_{+}\approx 0.144$ for $a_{2}\approx -0.084$ and the local minimum $k_{-}\approx 0.375$ for $a_{2}\approx -0.234$. For values of $p_0$ very close to $1$, i.e., for very low charged black holes, the horizons at $\theta=\pi/2$ may be very flattened. 

\subsection{The maximal proper time of free fall from the outer to the inner horizon}

Let us now consider the effect of the distortion fields on the black hole interior between its horizons. In particular, we would like to see if the inner horizon "can come any closer" to the outer one. In order to do so, we calculate the maximal proper time $\tau$ of free fall of a test particle from the outer to the inner horizon along the symmetry "semi-axes" $\theta=0$ and $\theta=\pi$. In this case, the maximal proper time of free fall corresponds to a time-like geodesic of zero azimuthal angular momenta associated with the Killing vectors $\xi_{(\phi)}^\mu$ and $\xi_{(\chi)}^\mu$ and zero energy, which is associated with the Killing vector $\xi_{(t)}^\mu$. For this free fall the coordinates $(t,\theta,\chi,\phi)$ remain constant along such geodesic. Thus, the proper time is a function of the time-like coordinate $\eta$, which changes from $1$ to $-1$. It is convenient to introduce another coordinate $\psi$ as
\be\n{7.1}
\eta=\cos\psi\hh \psi\in[0,\pi]\,.
\ee
Then, using the metric \eq{4.13}--\eq{4.15} together with the conditions \eq{5.3}, \eq{6.1}, and \eq{6.2} we derive
\ba
\hspace{-1.0cm}&&\tau_{|\theta=0}=
\sqrt{\frac{m_0}{8}}
\int_0^\pi d\psi\left[(1+p_0)(1+\cos\psi)e^{-2\tilde{u}_+(\psi)}
+(1-p_0)(1-\cos\psi)
e^{2\tilde{w}_+(\psi)}\right]^{\frac{1}{2}}\,,
\n{7.2}\\
\hspace{-1.0cm}&&\tau_{|\theta=\pi}=
\sqrt{\frac{m_0}{8}}
\int_0^\pi d\psi\left[(1+p_0)(1+\cos\psi)e^{-2\tilde{w}_-(\psi)}
+(1-p_0)(1-\cos\psi)
e^{2\tilde{u}_-(\psi)}\right]^{\frac{1}{2}}\,,
\n{7.3}
\ea
where we defined
\ba 
\tilde{u}_\pm(\psi)&=&
\sum_{n\geq0}(\pm 1)^na_{n}\cos^n\psi-u_0-\frac{1}{3}
(5u_1-4w_1)\,,\n{7.4}\\
\tilde{w}_\pm(\psi)&=&
\sum_{n\geq0}(\pm 1)^nb_{n}\cos^n\psi-w_0+\frac{1}{3}
(4u_1-5w_1)\,.\n{7.5}
\ea
Note that the expressions \eq{7.2} and \eq{7.3} are related to each other by the duality transformation \eq{5.16a}, \eq{5.18} accompanied by the transformation of the dummy variable $\psi\longrightarrow\pi-\psi$. 
\begin{figure}[htb]
\begin{center}
\hspace{0cm}
\ba
&&\includegraphics[width=6.5cm]{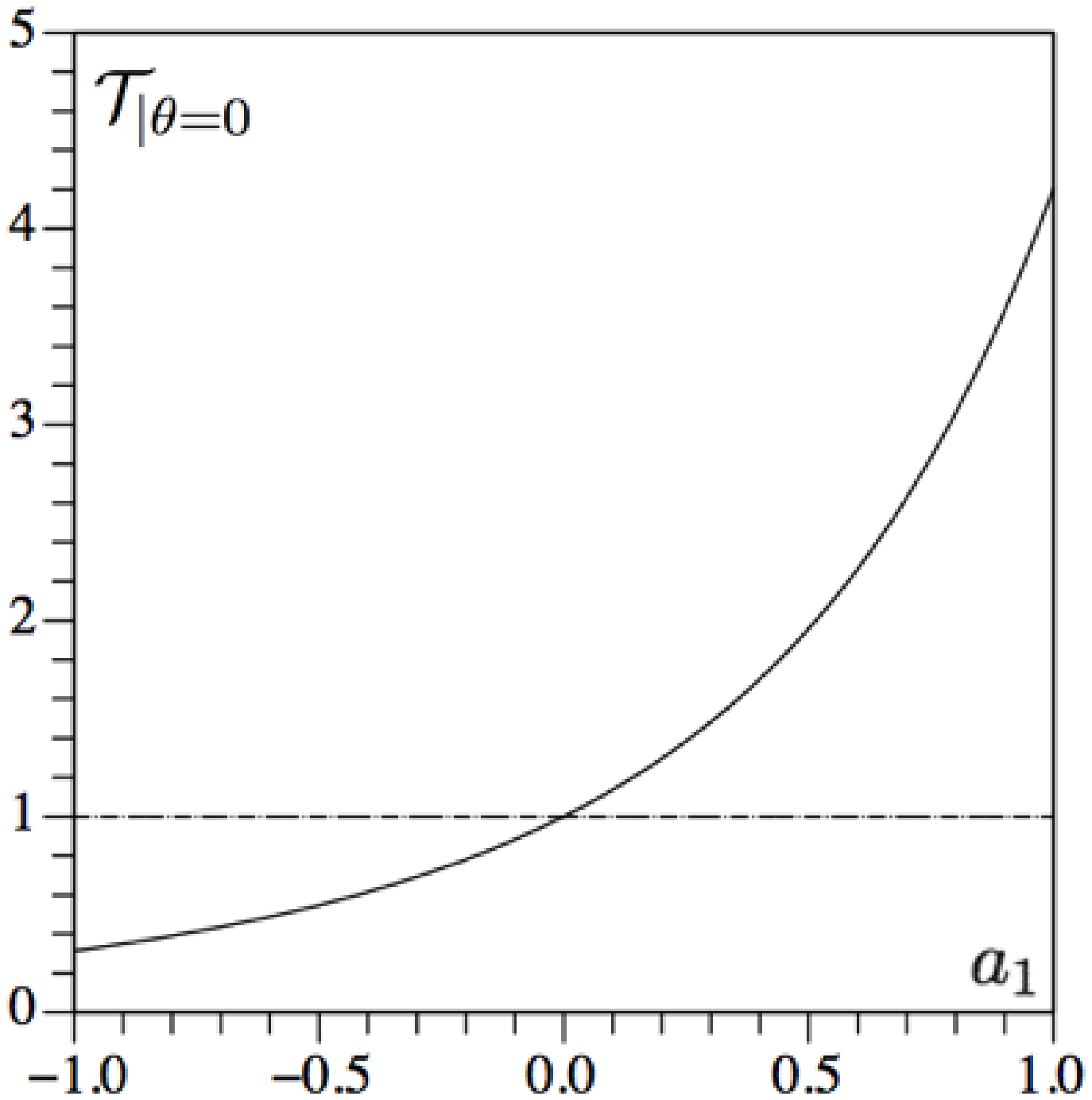}\hspace{0.7cm}
\includegraphics[width=6.5cm]{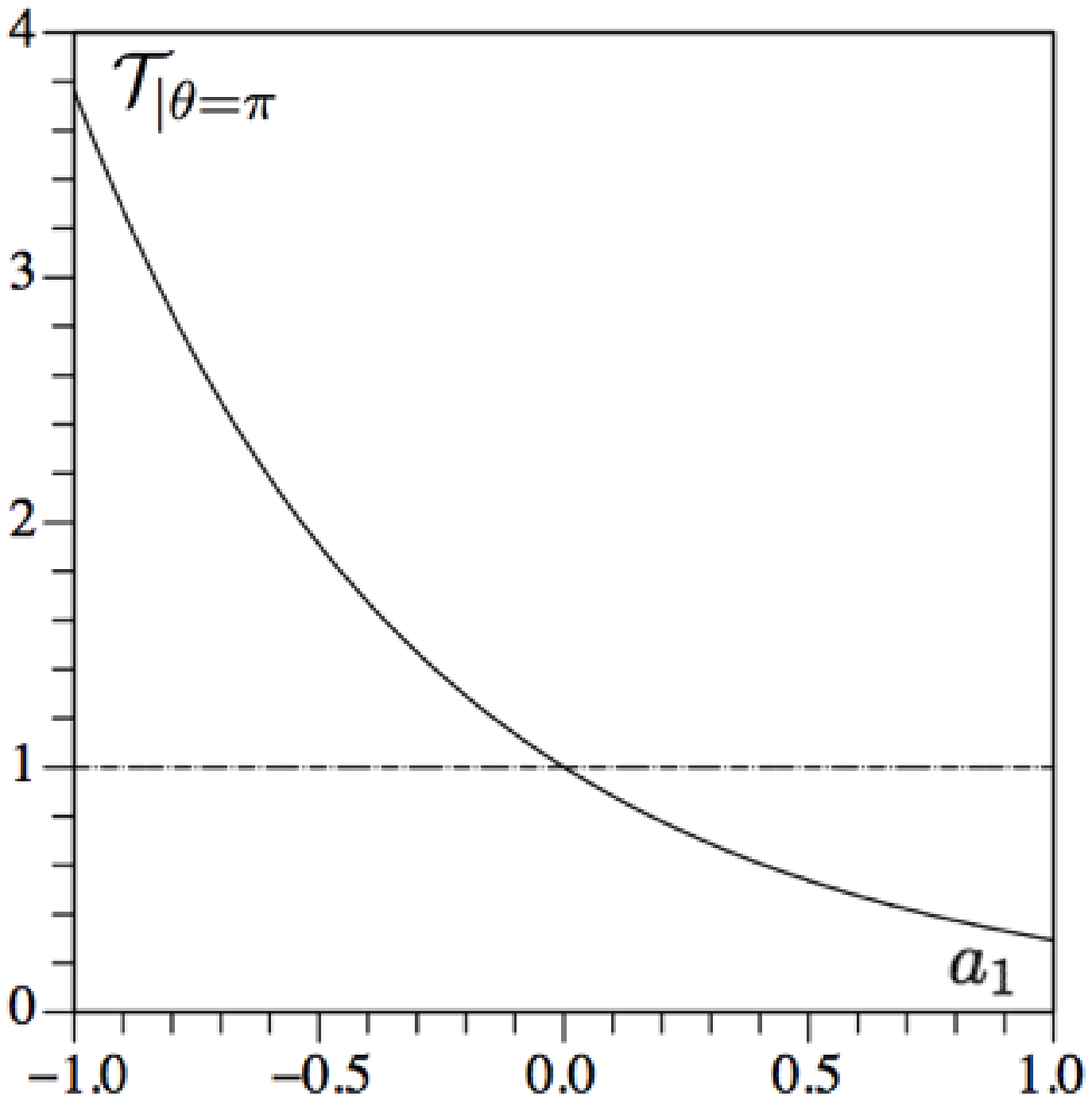}\non\\
&&\hspace{3.2cm}({\bf a})\hspace{6.5cm}({\bf b})\non
\ea
\caption{The maximal proper times for $p_{0}=0.6$ as functions of the dipole moment $a_1$ for free fall from the outer horizon to the inner horizon along the "semi-axes": ({\bf a}) $\theta=0$ and ({\bf b}) $\theta=\pi$. The horizontal dashed lines illustrate the maximal proper time corresponding to the Reissner-Nordstr\"om black hole.}\label{PTF5}
\end{center}
\end{figure}
\begin{figure}[htb]
\begin{center}
\hspace{0cm}
\ba
&&\includegraphics[width=6.7cm]{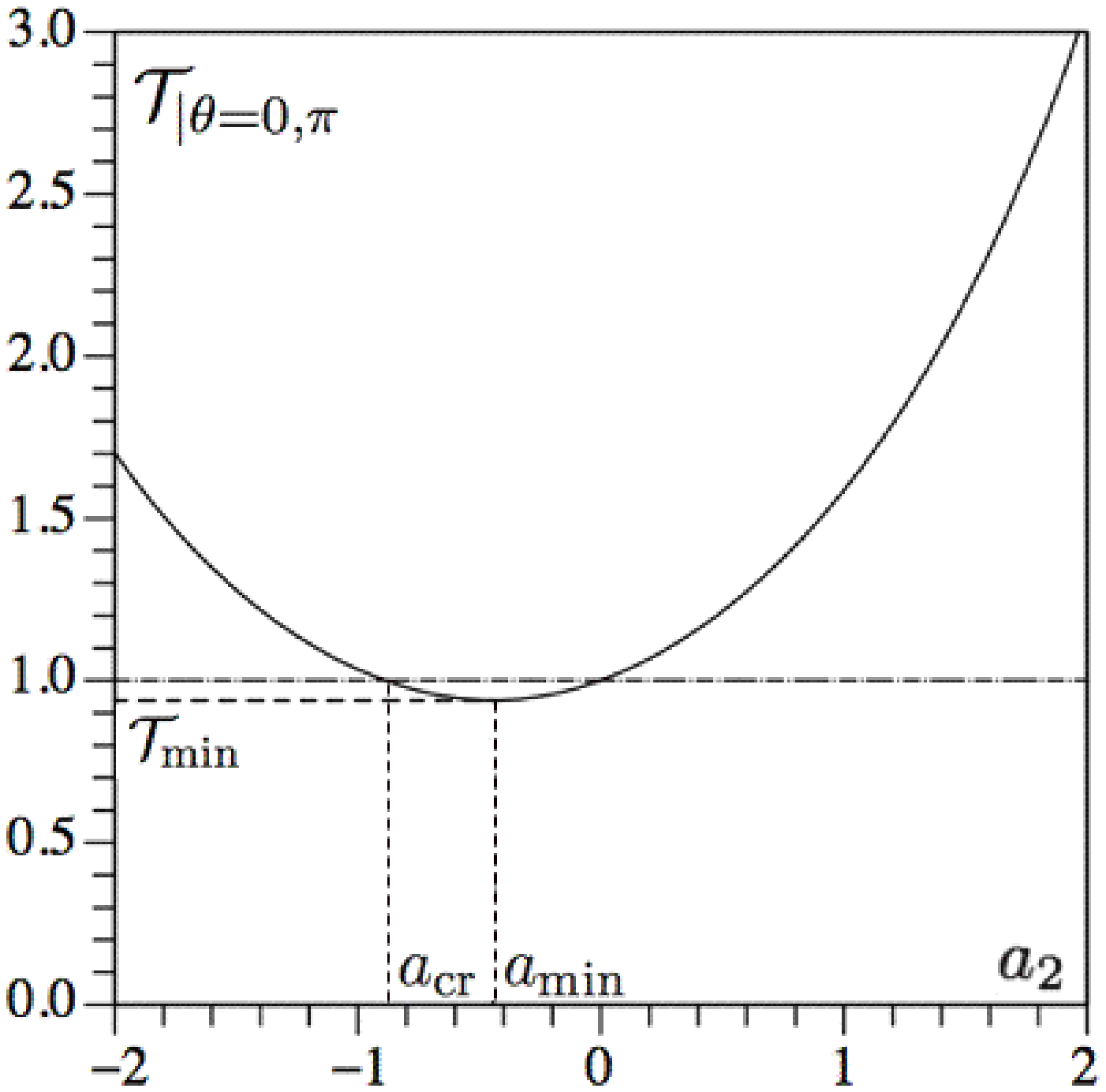}\hspace{0.7cm}
\includegraphics[width=6.5cm]{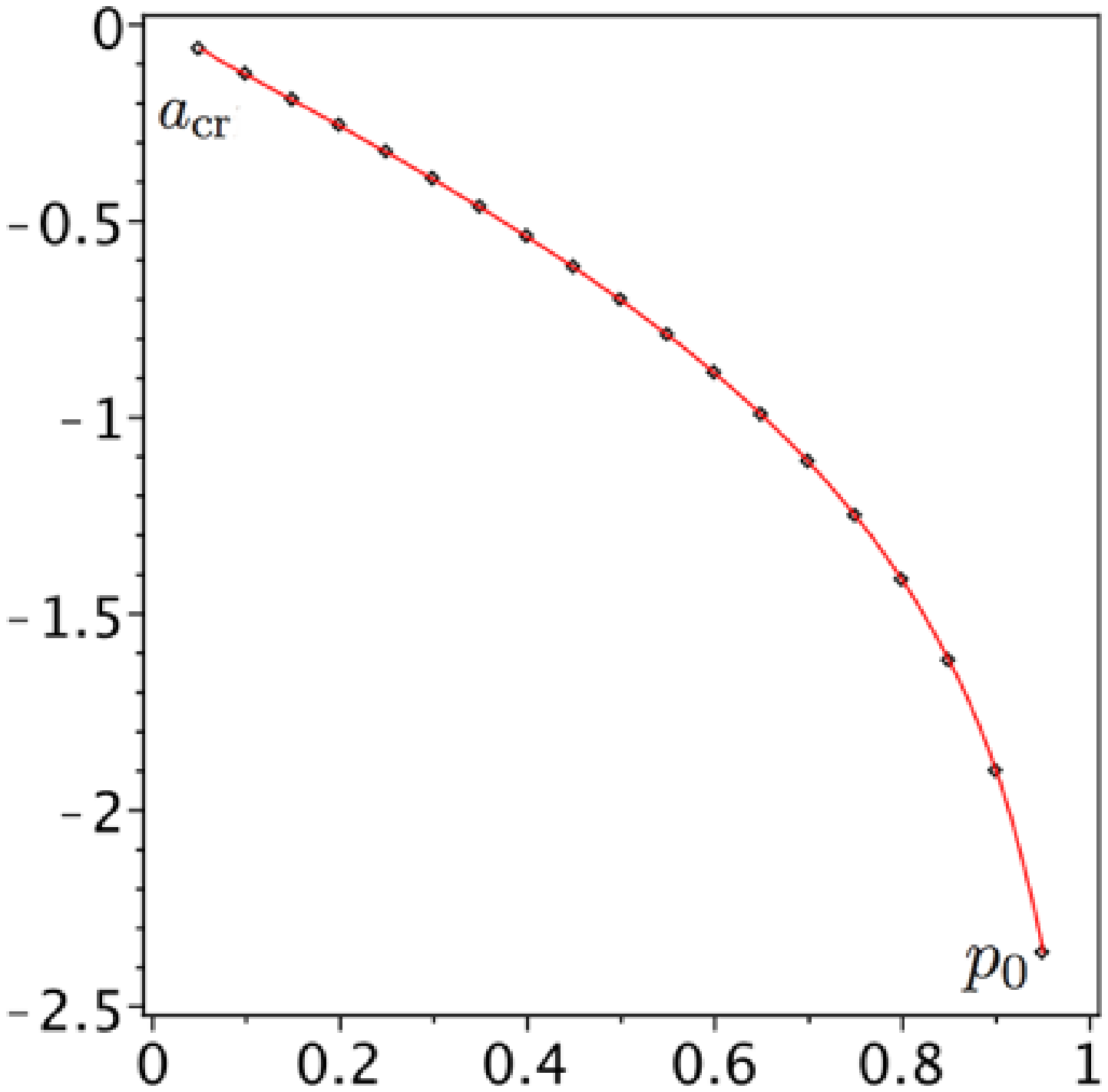}\non\\
&&\hspace{3.2cm}({\bf a})\hspace{6.7cm}({\bf b})\non
\ea
\caption{Plot ({\bf a}): The maximal proper times as functions of the quadrupole moment $a_2$ for free fall from the outer horizon to the inner horizon along the "semi-axes" $\theta=0$ and $\theta=\pi$ for $p_{0}=0.6$. The horizontal dashed line illustrates the maximal proper time corresponding to the Reissner-Nordstr\"om black hole. Plot ({\bf b}): The critical value $a_{\text{cr}}$ versus $0<p_{0}<1$. Note that for $a_{\text{cr}}<a_2<0$ the maximal proper time of free fall is less than that corresponding to the Reissner-Nordstr\"om black hole.}\label{PTF6}
\end{center}
\end{figure}
\begin{figure}[htb]
\begin{center}
\hspace{0cm}
\ba
&&\includegraphics[width=6.5cm]{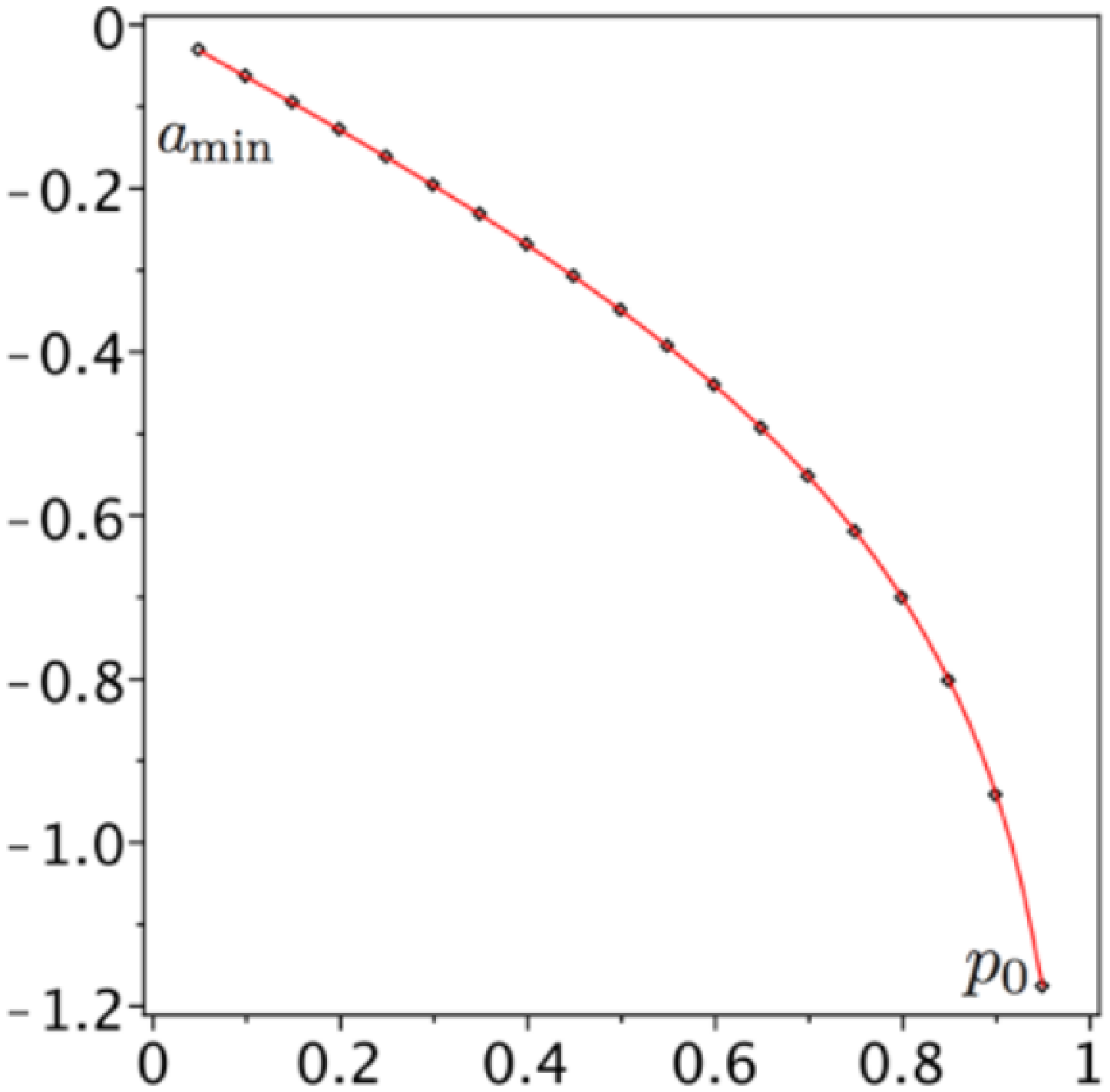}\hspace{0.7cm}
\includegraphics[width=6.5cm]{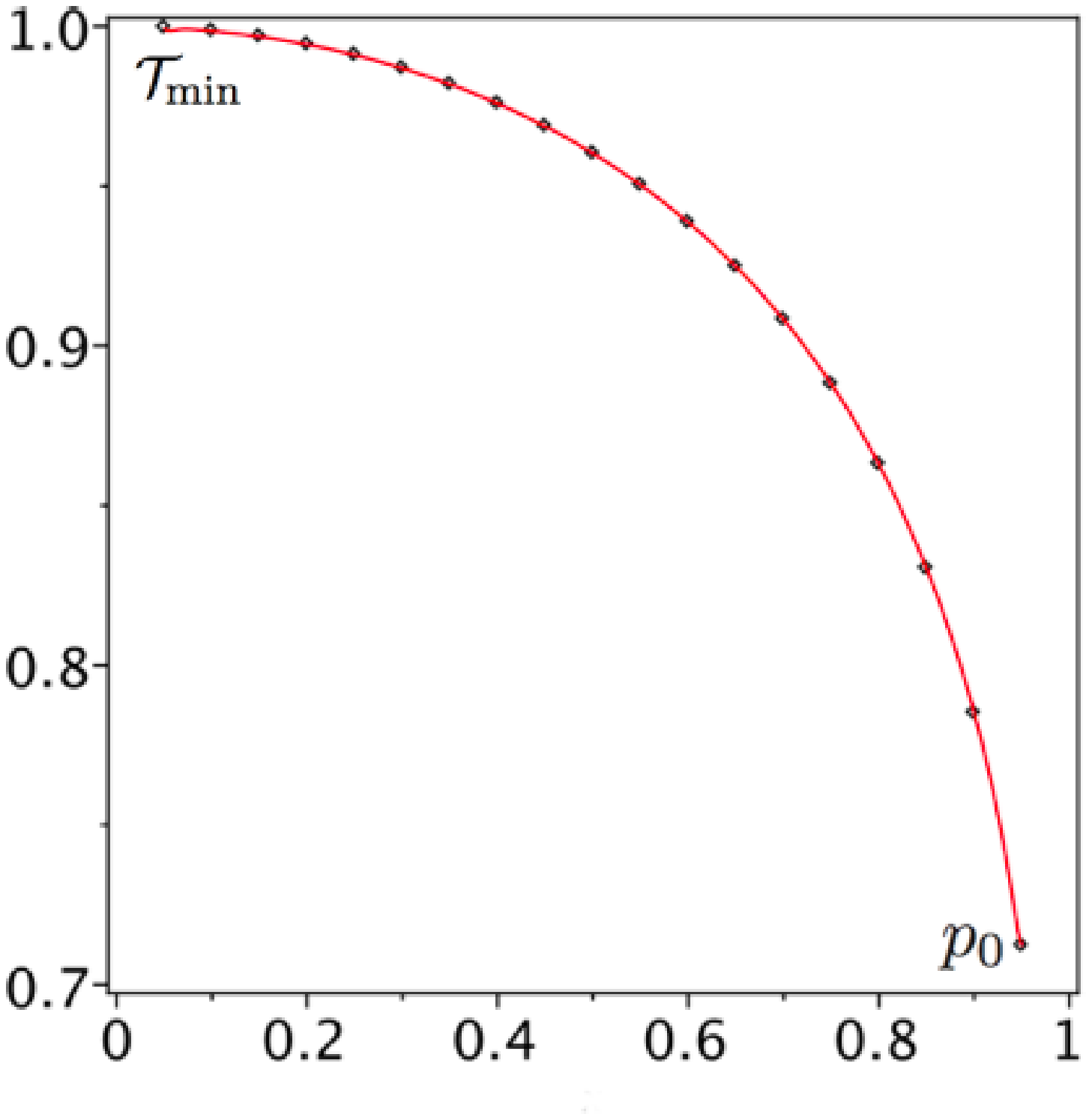}\non\\
&&\hspace{3.2cm}({\bf a})\hspace{6.7cm}({\bf b})\non
\ea
\caption{Plot ({\bf a}): The value of the quadrupole moment $a_{\text{min}}$ versus $0<p_{0}<1$. Plot ({\bf b}): The minimal value of $\mathcal{T}_{\text{min}}$ versus $0<p_{0}<1$.}\label{PTF7}
\end{center}
\end{figure}
We shall calculate the proper times $\tau_{|\theta=0}$ and $\tau_{|\theta=\pi}$ in units of the maximal proper time $\tau_{\text{RN}}$ corresponding to free fall from the outer to the inner horizon of the Reissner-Nordstr\"om black hole, which is
\be\n{7.6}
\tau_{\text{RN}}=\frac{\sqrt{m_0}}{2}
\int_0^\pi d\psi\sqrt{1+p_0\cos\psi}=
\sqrt{m_0(1+p_0)}E\left(\sqrt{\frac{2p_0}{1+p_0}}\right)\,,
\ee
where $E(x)$ is the complete elliptic integral of the second kind \cite{Abram}. Accordingly, we define the following quantities:
\be\n{7.7}
\mathcal{T}_{|\theta=0}=
\frac{\tau_{|\theta=0}}
{\tau_{\text{RN}}}\hh
\mathcal{T}_{|\theta=\pi}=
\frac{\tau_{|\theta=\pi}}
{\tau_{\text{RN}}}\,.
\ee

Let us now illustrate the effect of the dipole-monopole and quadrupole-quadrupole distortion fields on the maximal proper time of free fall. For the dipole-monopole distortion \eq{6.8} we have
\be\n{7.8}
\tilde{u}_\pm(\psi)=\pm a_1\cos\psi-\frac{5}{3}a_1\hh \tilde{w}_\pm(\psi)=\frac{4}{3}a_1\,,
\ee
and for the quadrupole-quadrupole distortion \eq{6.11} we have 
\be\n{7.9}
\tilde{u}_\pm(\psi)=\tilde{w}_\pm(\psi)
=-a_2\sin^2\psi\,.
\ee

The maximal proper times \eq{7.7} for $p_{0}=0.6$ as functions of the dipole moment $a_1$ are plotted in Fig.~\ref{PTF5}. From this Figure we see that for negative (positive) values of $a_{1}$ the maximal proper time $\tau$ of free fall from the outer to the inner horizon along the symmetry "semi-axis" $\theta=0$ is less (greater) than that of the Reissner-Nordstr\"om black hole. For free fall along the symmetry "semi-axis"  $\theta=\pi$, for positive (negative) values of $a_{1}$, the maximal proper time is less (greater) than that of the Reissner-Nordstr\"om black hole. This behavior of the maximal proper time is generic for all values of $p_{0}\in(0,1)$. The maximal proper times \eq{7.7} as functions of the quadrupole moment $a_2$ for $p_{0}=0.6$ are plotted in Fig.~\ref{PTF6}(a). For all values of $a_{2}$ between zero and the critical one, $a_{\text{cr}}$, the proper times are less than those of the undistorted Reissner-Nordstr\"om black hole. The dependence of $a_{\text{cr}}$ on $p_{0}$ is illustrated in Fig.~\ref{PTF6}(b). For a given value of $p_{0}$ there exists the corresponding value of $a_{\text{min}}$ for which the maximal proper time of free fall has its minimal value $\mathcal{T}_{\text{min}}$. Figure~\ref{PTF7} shows $a_{\text{min}}$ and $\mathcal{T}_{\text{min}}$ as functions of $p_{0}$. In contrast to the quadrupole-quadrupole distortion, the maximal proper time corresponding to the dipole-monopole distortion can be reduced to arbitrarily small values by taking arbitrarily large (but finite) negative (for $\theta=0$) or positive (for $\theta=\pi$) values of the dipole moment $a_{1}$.\footnote{One should take into account that for $|a_{1}|\gg1$ the higher order multipole moments cannot generally be neglected.} 

Summarizing our results, we see that the distortion fields can indeed bring closer to each other or move away the black hole horizons. In the case of the dipole-monopole distortion, depending on the values of the dipole moment $a_1$, the horizons can come arbitrarily close to or far from each other. While in the case of the quadrupole-quadrupole distortion, there is the range of the quadrupole moment $a_2\in(a_{\text{cr}},0)$ for which the horizons come close to each other and there is the minimal value $\mathcal{T}_{\text{min}}$ corresponding to $a_2=a_{\text{min}}$ which defines how close the horizons can come. The values $a_{\text{cr}},\mathcal{T}_{\text{min}}$, and $a_{\text{min}}$ decrease with increasing of $p_0$ [see Figs.~\ref{PTF6}({\bf b}), \ref{PTF7}({\bf a}) and \ref{PTF7}({\bf b})], i.e., with decreasing of the black hole's electric charge-to-mass ratio. In particular, for $p_0=0.05$ we have $q_0/m_0\approx 0.999$, and $\mathcal{T}_{\text{min}}\approx 0.999$, while for $p_0=0.95$ we have $q_0/m_0\approx 0.312$, and $\mathcal{T}_{\text{min}}\approx 0.712$.

\section{Conclusion}

In this paper, we studied distorted, five-dimensional, electrically charged (non-extremal) black holes on the example of a static and "axisymmetric" black hole distorted by external, electrically neutral, static and "axisymmetric" sources of gravitational field. The solution to the five-dimensional Einstein-Maxwell equations representing such a black hole was constructed by means of the procedure based on the gauge transformation of the matrix which is an element of the coset target space $SL(2,\mathbb{R})/U(1)$ of the scalar fields which define our model. In order to derive the solution, we applied the transformation to the vacuum seed solution representing distorted, five-dimensional, static and "axisymmetric", vacuum black hole. Note that taking the limit of vanishing electric charge, our solution becomes identical to the seed solution. The external sources of  the distortion are not included into the solution. As a result, the solution is not asymptotically flat. In fact, it diverges at the asymptotic infinity, where the sources are located. 
The space-time can be extended to achieve asymptotic flatness by including the sources into the solution. The constructed solution has the following properties:
\begin{itemize}
\item The space-time singularities (beside those corresponding to the location of the sources) are located behind the inner (Cauchy) horizon of the distorted black hole, provided that the sources of distortion fields satisfy the strong energy condition. 

\item There is a duality transformation between the outer and the inner horizons of the distorted black hole. This duality transformation is exactly the same as that between the horizon and the {\em stretched} singularity surfaces of the distorted, five-dimensional, static and "axisymmetric", vacuum black hole, which is the seed solution. The same case happens to be in four dimensions, where the duality transformation between the outer and the inner horizon surfaces of a distorted, electrically charged, static and axisymmetric black hole is exactly the same as that between the horizon and the {\em stretched} singularity surfaces of the corresponding distorted, electrically neutral black hole \cite{AFS.IV}. The duality transformation corresponds to an exchange between the symmetry "semi-axes" of our solution and reverse of signs of the multipole moments, which define the distortion fields.

\item We calculated the horizon surface areas and found that their product depends only on the black hole's electric charge and equals to the horizon area product of the Reissner-Nordstr\"om (undistorted) black hole of the same charge value. In addition, we constructed the universal area inequality which shows that the geometric mean of the areas is the upper (lower) limit for the inner (outer) horizon area. The area product and the universal area inequality are the higher-dimensional generalizations of those of a distorted four-dimensional black hole (see \cite{Hen1,Hen2,Hen3,Hennig1}). The calculated areas, surface gravity and the electrostatic potential at the black hole horizons satisfy the Smarr formula given for both the horizons, which is exactly the same as that for the five-dimensional Reissner-Nordstr\"om (undistorted) black hole. In addition, we found that the electromagnetic field invariant calculated at the black hole horizons is proportional to the squared surface gravity of the horizons. The coefficient of proportionality depends on the charge-to-mass ratio of the black hole and is independent of the distortion parameters. The quantities calculated at the black hole horizons are related by the duality transformation. 

\item The Kretschmann scalars calculated at the black hole horizons are related by the duality transformation, which implies that if the Kretschamnn scalar calculated at the outer horizon is regular, then the Kretschmann scalar calculated at the inner (Cauchy) horizon is regular as well. The Kretschmann scalars depend on the trace of the square of the Ricci tensor and the Ricci scalar of the horizon surfaces and on the electromagnetic field invariant calculated at the horizons, or the horizons surface gravity. These quantities are finite if the distortion fields are regular and smooth at the black hole horizons.

\item The outer horizon area of a black hole distorted adiabatically does not change. Because the product of the inner and outer horizon areas depends on the black hole's electric charge only, and the distortion fields do not change its value, the area of the inner horizon does not change under adiabatic distortion either. As a result, the expressions for the Kretschmann scalars calculated at the black hole horizons can be presented in the form independent of the monopole moments of the distortion fields. On the example of the dipole-monopole and quadrupole-quadrupole distortion fields we illustrated in the plots that, as a result of the distortion, both the inner and outer horizons have regions of high and low space-time curvature. For the given range of the dipole and quadrupole moments, the maximal values of the Kretschmann scalar calculated at the outer horizon in the units of that of the Reissner-Nordstr\"om (undistorted) black hole of the same mass and electric charge are 10~---~20 times greater than those of the Kretschmann scalar calculated at the inner horizon. It implies that the outer horizon is more susceptible to the distortion than the inner one. In the case of the quadrupole-quadrupole distortion, there are regions of the outer black hole horizon where the space-time curvature can be very low, especially if the black hole's charge-to-mass ratio is small.

\item The distortion fields affect noticeably the black hole interior region. Studying the maximal proper time of free fall of a test particle from the outer to the inner horizon along the symmetry "semi-axes" we showed on the example of the dipole-monopole and quadrupole-quadrupole distortion fields that due to the distortion the horizons can come close to each other or move away. In the case of the dipole-monopole distortion, depending on the value of the dipole moment, the horizons can either come arbitrarily close to or move far from each other, while in the case of the quadrupole-quadrupole distortion where is the minimal value of the horizons approach, which cannot be arbitrary small. This minimal value decreases with the decreasing value of the black hole's charge-to-mass ratio. These results show that the effect of the distortion fields on the black hole interior depends not only on their strength but as well on their type. 
\end{itemize}

In addition, we formulated the zeroth and the first law of mechanics and thermodynamics of the distorted black hole and found a correspondence between the global and the local forms of the first law. These laws are higher-dimensional generalizations of the laws formulated for four-dimensional, distorted, electrically changed black hole by Fairhurst and Krishnan \cite{Fai.IV}, who considered more general class of distortion fields generated by electrically charged matter.

There are some open issues which we would like to mention here. The established  duality transformation illustrates a correlation between the space-time geometry at the black hole horizons. Such a transformation seems to be an inherent property of a Weyl-type solution. In the case of arbitrarily distorted, four-dimensional, electrically charged, stationary and axisymmetric black hole, the corresponding Einstein-Maxwell equations are equivalent to two complex Ernst equations. These equations can be viewed as the integrability conditions for an associated linear problem (see \cite{Hen2,Hen3}). Then, as it was demonstrated by Ansorg and Hennig \cite{Hen2,Hen3}), the duality transformation (in the general form) emerges from the integration of the linear problem along the black hole horizons and the symmetry axis. Such a linear problem can be formulated by means of a Lax pair construction for more general, higher-dimensional models of gravity (see, e.g., \cite{Gal,GR}). Existence of a Lax pair is directly related to a generation of an infinite number of solutions, starting from a known one, and thus, to a complete integrability (see, e.g., \cite{Bel1,Bel2}). Thus, one may expect that a certain duality transformation may exist between horizons of some solutions of these models, which possess a certain group of isometries. 

As we illustrated on the example of the five-dimensional, electrically charged, static and "axisymmetric"  black hole, distorted by electrically neutral matter, its inner (Cauchy) horizon remains regular. It is rather possible (because of the duality transformation) that it will be regular under more general type of distortion due to electrically charged matter. However, whether the inner horizon is regular under arbitrary (asymmetric) static distortion, is an open question in four- and five-dimensional space-times.

One natural generalization of the distorted, five-dimensional, static black hole is to consider a distorted, five-dimensional, stationary black hole. The analysis presented here can be done on other distorted, higher-dimensional black objects, such as black strings, black rings, black saturns, etc.

\acknowledgments

S. A. gratefully acknowledges the Deutsche Forschungsgemeinschaft (DFG) for financial support within the framework of the DFG Research Training group 1620 Models of gravity. A. A. S. is grateful to the Natural Sciences and Engineering Research Council of Canada for its financial support.

\end{document}